\documentclass[aps,prb,superscriptaddress,twocolumn]{revtex4-2}

\usepackage{setspace}
\usepackage{lmodern}
\usepackage[utf8]{inputenc}

\usepackage{amsmath}
\usepackage{amssymb}
\usepackage{graphicx}
\usepackage{epstopdf}
\usepackage{color}
\usepackage{enumerate}
\usepackage{bm} 
\usepackage{physics}

\usepackage[dvipsnames]{xcolor}

\usepackage[labelfont=bf]{caption}
\usepackage{multirow}
\usepackage{floatpag}
\usepackage[misc]{ifsym}
\usepackage{sistyle}
\usepackage{upgreek}
\usepackage{upgreek}
\usepackage{hyperref}

\usepackage{dsfont}
\usepackage{array}
\usepackage{physics}
\usepackage[normalem]{ulem}
\usepackage{upgreek}
\usepackage{xcolor}
\usepackage{hyperref}
\usepackage{comment}
\usepackage{enumitem}
\usepackage{soul}
\usepackage[capitalize]{cleveref}
\usepackage{quantikz}

\usepackage{refcount} 

\usepackage[framemethod=default]{mdframed}

\definecolor{dark-red}{rgb}{0.84,0.15,0.15}
\definecolor{dark-blue}{rgb}{0.15,0.15,0.4}
\definecolor{medium-blue}{rgb}{0,0,0.5}
\definecolor{copper}{rgb}{0.72, 0.45, 0.2}

\hypersetup{
    colorlinks, linkcolor={dark-red},
    citecolor={dark-blue}, urlcolor={medium-blue}
}\definecolor{dark-red}{rgb}{0.84,0.15,0.15}

\begin{document}

% \title{Hilbert space signatures of non-ergodic glassy dynamics}
\title{Multifractal and Glassy Signatures of Non-Ergodic 2D Quantum Dynamics}

\author{Google Quantum AI and collaborators}

\begin{abstract}
The fate of ergodicity and localization in disordered, interacting quantum systems in two spatial dimensions remains an outstanding open problem in non-equilibrium physics~\cite{BAA2006,rev2,rev3,rev01,rev4,gopalakrishnan2019instability,rev1,suntajs2020quantum,sels2021dynamical,peacock2023manybody,rev5,de2013ergodicity}. Theoretical and numerical approaches are severely constrained by exponential Hilbert-space growth. Here, we leverage the high data-acquisition rates of superconducting quantum processors to extensively sample Hilbert-space configurations, effectively imaging the state as a wavefunction network (WFN)~\cite{Network_PRX2024,andreoni2025network}. Tracking the bitstring return probability $R(t)$ across varied disorder strengths, we find heavy-tailed distributions and power-law typical decay characteristic of glassy relaxation. Furthermore, the distribution of bitstring probabilities transitions from a Porter--Thomas form at low disorder to an eight-decade power law at stronger disorder. In the reconstructed configuration space, WFN connectivity is compatible with a random geometric network at low disorder and becomes scale-free at stronger disorder. Finally, a multifractal analysis based on the binary intrinsic dimension (BID) reveals a non-ergodic regime where the wavefunction support saturates to a finite fraction of the total Hilbert space. Together, these results rule out a direct ergodic-to-localized transition, pointing instead to an extended multifractal regime. By visualizing wavefunction dynamics directly in configuration space, this work complements traditional real-space studies, showing the potential of quantum processors to resolve long-standing questions in non-equilibrium physics.
\end{abstract}

\date{\today}
\maketitle 

To date, most quantum simulation experiments have served to verify established theoretical and numerical predictions. Moving beyond verification requires leveraging quantum processors to measure many-body quantities that have remained inaccessible to conventional probes. Paradigmatic examples include scanning tunneling microscopy and angle-resolved photoemission spectroscopy, which unlocked new many-body phenomena by directly resolving real- and momentum-space structures. Here, we extend this paradigm to Hilbert space, introducing an experimental approach that directly resolves configuration-space dynamics, moving beyond conventional low-order correlation descriptions.

Diffusion and localization of many-body particles are often theoretically formulated as hopping processes in abstract spaces, a framework established by early works on configuration-space graphs and later adopted in the many-body localization\,(MBL) studies, where localization is analyzed in Fock space~\cite{AAT,Logan1990,BAA2006,altshuler1997quasiparticle}.\,To date, most experimental studies of ergodicity focus on real-space observables, such as relaxation rates and the asymptotic behavior of spatial correlations\,\cite{schreiber2015observation, choi2016, bordia2017periodically, li2025many, hur2025stability}.\,Explicit experimental visualization of dynamics in configuration space has remained elusive; requiring repeated sampling of wavefunctions in the configuration space basis. With data-taking rates reaching 100\,kHz to 1\,MHz, quantum processors now meet these sampling demands\,\cite{kjaergaard2020superconducting,google2025quantum}.

\vspace{-1mm}
\begin{figure}[th!]
    \centering
    \includegraphics[width=0.41\textwidth]{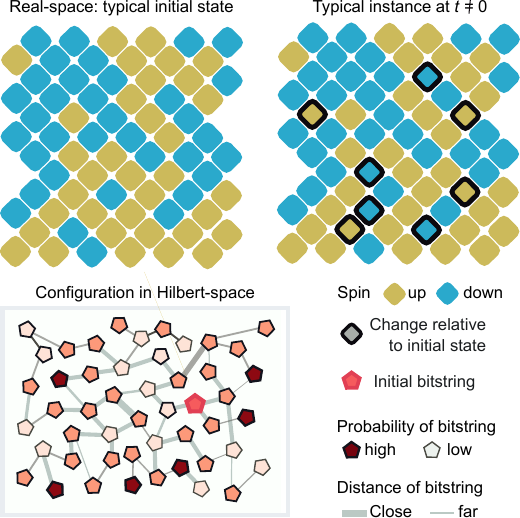}
    
\caption {\textbf{Wavefunction networks.} A typical initial state of $n=70$ qubits alongside measured $z$-basis configurations (bitstrings) at time $t$. Extensively sampling such bitstring snapshots enables us to image Hilbert space by constructing the wavefunction network (WFN) in configuration space.}
    \label{fig:hilbert_space}
\end{figure}

\begin{figure*}[th!]
    \centering
    \includegraphics[width=\textwidth]{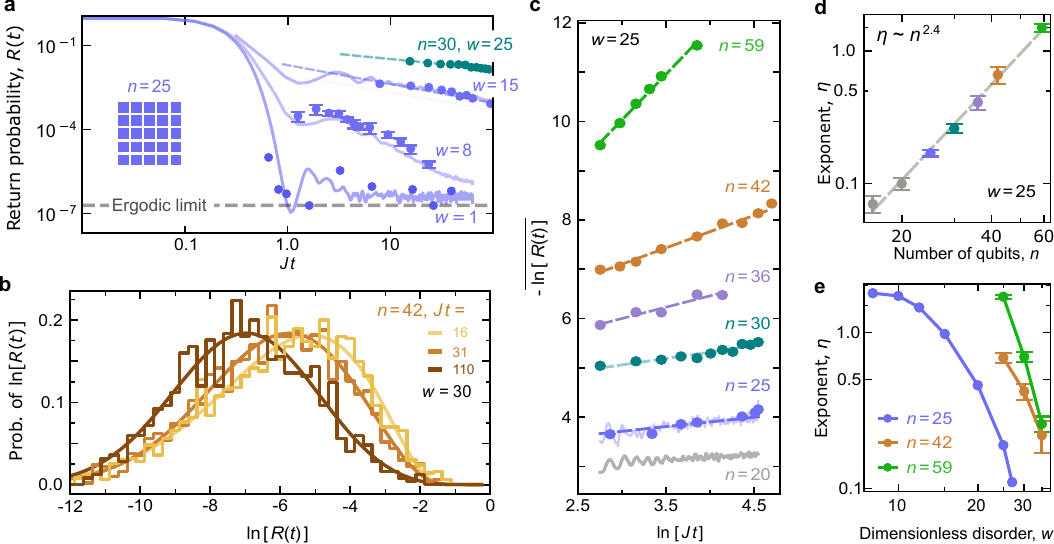}
    \caption{\textbf{Signatures of glassy behaviour in $R(t)$. a,} 
      $R(t)$ for $n=25$ (circles) at low \,$w$;  $R(t)$ shows a power-law decay for larger $w$. Solid curves are numerics averaged over $\sim 2500$ disorder instances (100 for $w=1$). Straight lines indicate power-law fits.\,\textbf{b,} R(t) for $n=42$ on logarithmic scales at three different evolution times for $w=30$. Solid lines are Gaussian fits.\,\textbf{c,} $R(t)$ for $w=25$ and for a range of system sizes. Solid straight lines show power-law fits.\,\textbf{d,} The power-law exponents extracted in \textbf{c} scale as $\eta\propto n^{2.4}$.\,\textbf{e}, Exponent $\eta$ as a function of disorder for different system sizes. Continuous lines in \textbf{c,} $n=20,25$, and gray symbols in \textbf{d,} $n=16, 20$ are numerics; for $n=25$ numerical results coincide with experimental ones.} \label{R(t)}
\end{figure*}

\vspace{0.8mm}
Directly probing configuration space offers a fundamental experimental advantage. In real space, resolving quantities like diffusivity is limited by the system's smallest linear dimension, scaling as $\sim \sqrt{n}$ for a 2D lattice of $n$ spins; this obscures, for example, power law behavior that necessarily spans several orders of magnitude. In the time domain, tracking quantities like population imbalance suffers from a poor signal-to-noise ratio. Conversely, the exponential growth of Hilbert space with system size ($N = 2^n$) supports a vast dynamic range. While many snapshots are necessary to resolve processes spanning this expanded range, in disordered systems, the required number is far lower than the naive $\mathcal{O}(2^n)$ prediction; this is in part due to the fact that the questions asked are focused on the shape or dimensionality of wavefunctions. As we will show, these geometric quantities can typically be resolved with manageable $10^5$--$10^8$ shots per disorder realization.

The dynamics of excitations in a 2D nearest-neighbor qubit array can be described by the spin Hamiltonian
\begin{equation}
H = -J \sum_{\langle i, j \rangle} (S^+_i S^-_j + S^-_i S^+_j ) + \sum_i h_i S_i^z\,, \label{H1}
\end{equation}
\noindent where \( S_i^\alpha \) are spin-\(\tfrac{1}{2}\) operators, \( h_i \) are site-dependent random disorder drawn from a box distribution of width \( W \) and zero mean, and \( J \) denotes the nearest-neighbor coupling strength. We prepare an initial state in the configuration ($S_z$) basis, evolve the state for time $t$, and then measure in the configuration basis. Collecting statistics of final state `snapshots' as a function of $t$ allows us to reconstruct a Hilbert-space view of the dynamics. We report results for three quantities: (i) the return probability\,(Fig.~\ref{R(t)}); (ii) the distribution of observed configurations\,(Fig.~\ref{WavefuncStats}); and (iii) the structure of the wavefunction as a configuration-space graph\,(Fig.~\ref{Network},\ref{Graph_analysis}). In the SI, we present, for comparison, related real-space observables.

\begin{figure*}[t!]
\includegraphics[width=0.98\textwidth]{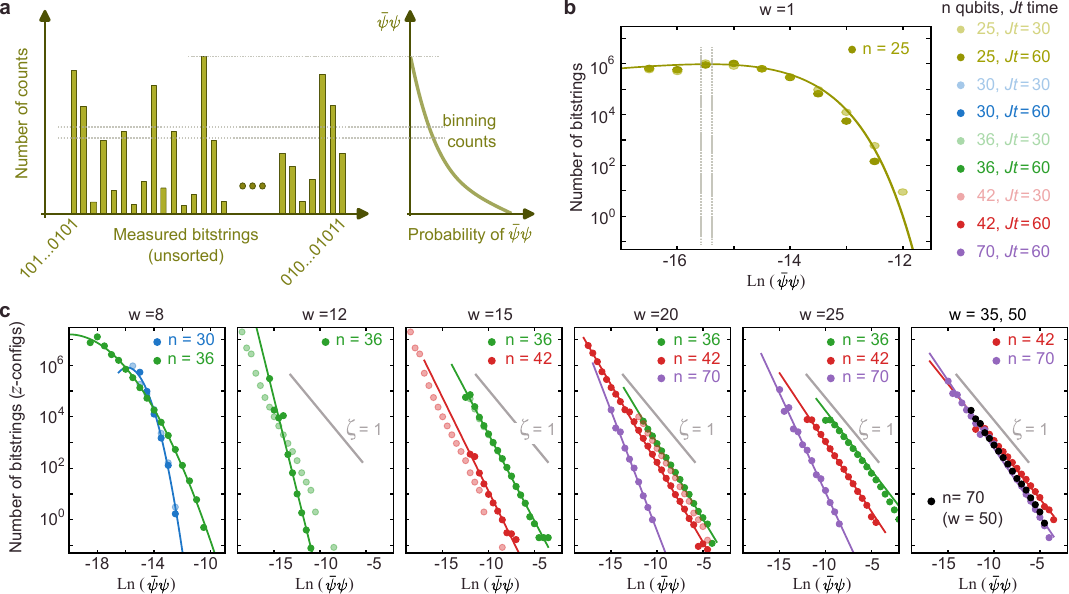}
\centering
\caption{\textbf{Wavefunction statistics. a,} Repeated measurements of bitstrings allows us to associate a Born probability, $\bar{\psi}\psi$, to each observed $z$-config.\,\textbf{b,} Distribution of counts vs. $\ln(\bar{\psi}\psi)$ for $n=25$ at $w=1$; solid line fit to the Porter-Thomas distribution.\,\textbf{c,} For $w=8$, the data are fit to $\mathcal{P}(u) du$ with $u = \ln(x)$ and $\mathcal{P}(u) \propto \exp(-a (u- u_m)^2)$, with fitting parameters $(a,u_m)_{n=30} = (1.1,-3.0) $ and $(a,u_m)_{n=36} = (0.185,-2.9) $. For $w\geq 12$, all data are fit to $\mathcal{P}(u)\sim e^{-\zeta u} $. Solid and faded dots correspond to $Jt=30$ and $Jt=60$, respectively.
%\,\textbf{d,} The height of each bar is the total number of observed bitstrings at a fixed Hamming distance relative to most likely $z$-config; the color of each point is associated with the probability of bitstrings.
} \label{WavefuncStats}
\end{figure*}

\begin{figure*}[t!]
\includegraphics[width=\textwidth]{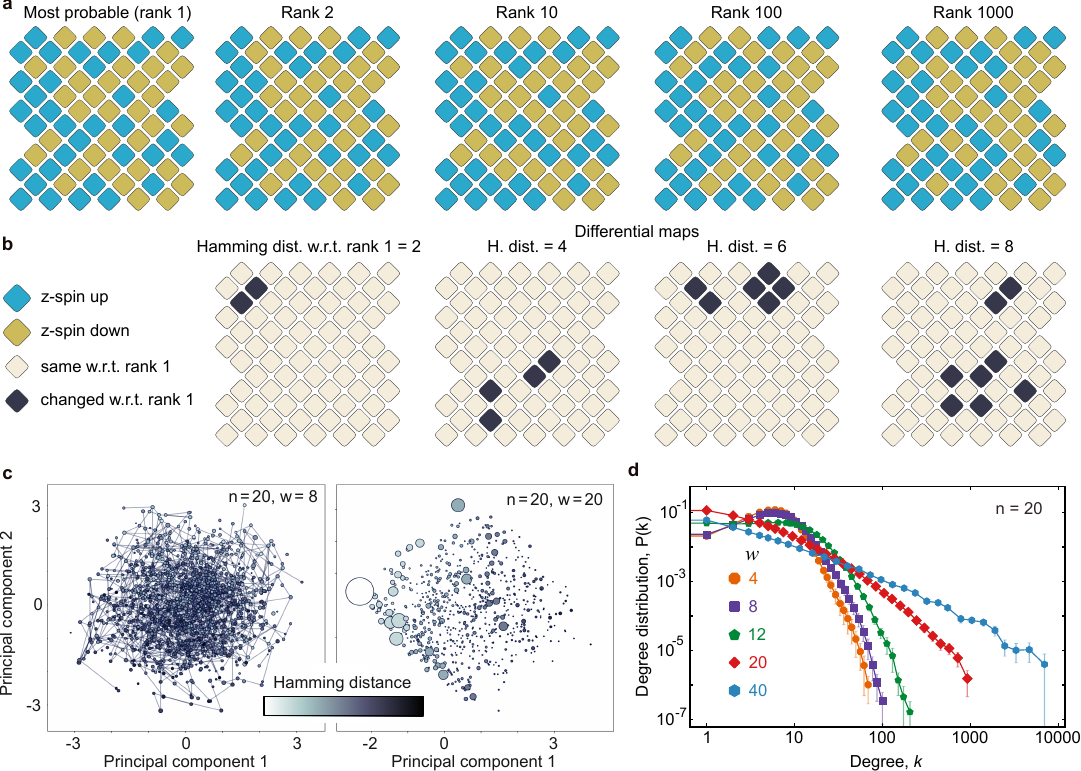}
\centering
\caption{\textbf{Constructing Wavefunction networks\,(WFN). a,} The most probable (rank 1) and four other less-probable bitstrings for a given disorder realization of strength $w=35$ at $Jt=60$.\,\textbf{b}, Bit-flip maps showing the difference relative to the most probable bit-string.
\,\textbf{c}, Representation of the data as a WFN for a disorder realization instance with $n=20$ and $w=8,20$. States in configuration space are encoded as nodes; links are drawn geometrically within a cutoff defined as the average 5th-nearest neighbor Hamming distance. Size is proportional to the node's connectivity, and color indicates Hamming distance from the initial bitstring. The graphs are projected onto the first two principal axes derived via Principal Component Analysis (PCA).\,\textbf{d}, Normalized degree distribution $P(k)$ of the data for $n=20$ represented as WFNs, built as in panel c, for various values of the disorder; $k$ denotes the number of direct network connections per $z$-config. Computed using a representative sample of 10,000 bitstrings, with errorbars obtained from averaging over 10 disorder realizations for 10 initial bitstrings.} \label{Network}
\end{figure*}

% \,\textbf{c}, The top $100 \times 100$ corner of an adjacency matrix for a system of size $n=70$, representing the mutual hamming distances between the 100 most probable bitstrings. \,\textbf{d}, Graph representation of wavefunction spreading for $n=70$ at a single disorder instance $w=35$. States in configuration space are encoded as nodes; size is proportional to $|\psi|^2$, and the width of an edge is proportional to the Hamming distance separating the two nodes under consideration.

The return probability to initial state $| \Psi(0)\rangle$, 
\begin{equation}
R(t) = |\langle \Psi(t)| \Psi(0) \rangle|^2\,,
\end{equation}
\noindent is a direct measure of ergodicity. In an ergodic system, $R(t)$ should quickly drop to the inverse volume of the total Hilbert space. In the ergodic regime, $w\sim 1$, $R(t)$ decays quickly to this value\,(Fig.\,\ref{R(t)}a). Before asymptoting, however, $R(t)$ shows a shallow minimum around $Jt\sim1$, consistent with a ``correlation hole"~\cite{Leviandier86,Tores2018}. At larger disorder strengths, $w>10$, the correlation hole disappears and $R(t)$ begins to hint at power-law dependence,
\begin{equation}
    R_{\rm{typ}}(t) \sim (Jt)^{-\eta}
    \label{R-power}
\end{equation}
\noindent Concomitantly, $R(t)$ at any given time acquires a very wide distribution, even on the logarithmic scale (Fig. \ref{R(t)}b). We fit the distribution of $\ln(R)$ with Gaussians to obtain the dependence of the typical $\overline{\ln R(t)}$ on system size, shown in Fig. \ref{R(t)}c. The decay of the typical return probability is consistent with a power law for all system sizes considered, and is also compatible with 1D numerical results and the ``Quantum Sun" model~\cite{Tores2015,Vidmar1,Vidmar2}. In classical glasses, tunneling between nearby energy wells generates low-frequency noise~\cite{CuKu, Rejuvenation,Esquinazi, Kogan, Noise}. The observed power-law decay of $R_{\rm{typ}}$ for $10 \leq w \leq 35$ in Fig.~\ref{R(t)} is reminiscent of such noise. This decay suggests that residual interactions lead to dephasing and inelastic transitions, where a logarithmic growth in spin-flip probability generates the $1/f$ noise characteristic of glasses and spin glasses\,(see Secs.~\ref{subsec:short_time} and~\ref{subsec:spin_noise} of the SI). A transition from superfluid to Bose glass state at $w \approx 19$ was found for the same model at zero temperature in Ref.~\cite{Laflorencie2015}.  Note that the exponent $\eta$ grows super-linearly with $n$,
$\eta \approx \kappa(w)\,\cdot n^{2.4}$, as plotted in Fig.~\ref{R(t)}d. At large $w$, $\eta$ decreases for all values of $n$ (Fig.\,\ref{R(t)}e), suggesting the appearance of more localized states at higher disorder strengths. Since the number of independent spin pairs scales linearly on $n$, the superlinear dependence of $\eta$ suggests that dephasing is dominated by internal many-body dynamics rather than external noise. As $n$ increases, the number of active spin pairs contributing to local dephasing processes at any given site eventually saturates. Consequently, $\eta$ is expected to transition back to linear scaling at large sizes ($n \geq w^2$), (see Secs.~\ref{subsec:exponent_eta} and~\ref{subsec:dephasing_without} of the SI). Note that $R(t)$ can also be thought of as an autocorrelation in Hilbert space; for comparison, we present results on the real-space spin-spin correlation in the SI. 

\begin{figure*}[th!]
    \centering
    \includegraphics[width=0.99\textwidth]{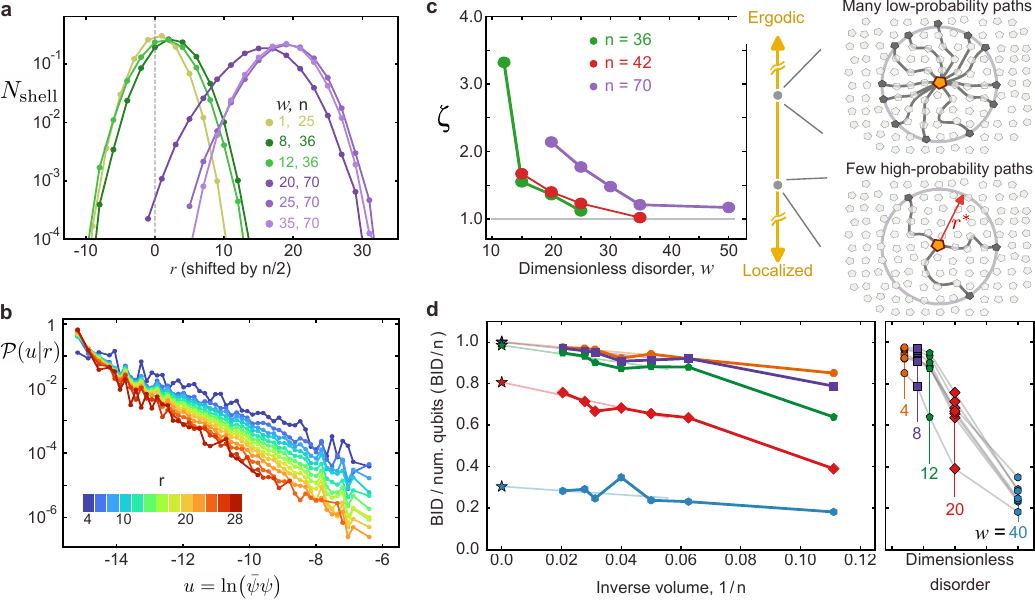}
    \caption{\textbf{Non-ergodicity and multifractality.} \,\textbf{a},\, Normalized shell density $N_\text{shell}(r)$, shifted by $r=13, 18, 35$ for $n=25, 36, 70$, respectively. \textbf{b},\, Marginal probability distributions over log-spaced probability bins, coloured by Hamming distance $r$, for a representative sample $n=70$, $w=35$. \textbf{c},\, Exponent $\zeta$ vs. $w$ for $n=36,42,70$.\,\textbf{d,}\, Normalized Binary Intrinsic dimension (BID) vs. inverse system size for various $w$ values, with linear extrapolations of the three biggest sizes to the thermodynamic limit ($1/n \rightarrow0$). Computed from 10,000 representative samples. The side panel shows the same data as a function of $w$ to emphasize the transition. Error-bars are obtained from averaging over 10 disorder realizations for 10 initial bitstrings.}
    \label{Graph_analysis}
\end{figure*}

Next we measure the probability distribution over final $z$-configurations (Fig.~\ref{WavefuncStats}).\,Fully ergodic wavefunctions obey the Porter-Thomas\,\cite{PorterThomas} distribution, $\mathcal{P}_\text{PT}(N|\psi|^2)$, where $\psi \equiv \psi[z]$ is a many-body wavefunction in the $z$-basis; $N$ is the Hilbert space dimension~\cite{N}; and $\mathcal{P}_\text{PT}(x) = \frac{1}{\sqrt{2\pi x}}e^{-x/2}$. In Fig.\,\ref{WavefuncStats}b we confirm that our data follow this distribution at low disorder, $w=1$. Increasing disorder qualitatively changes the fit. Close to $w=8$, the distribution broadens significantly and no longer obeys the Porter-Thomas form; instead it becomes log-normal, as shown in the left panel of Fig.\,\ref{WavefuncStats}c. Such distributions appear in the context of mesoscopic fluctuations and (weak) Anderson localization\,\cite{AKL1986,FalkoEfetov1995,MirlinEvers}.  At higher disorder, the distribution convincingly follows power-law behavior---spanning 8 orders of magnitude in some cases. These data are shown in Fig.\,\ref{WavefuncStats}c, fit to
\begin{equation}
    \mathcal{P}(x) \sim x^{-\zeta -1}.
    \label{Ppsi}
\end{equation} 
The middle panels\,($w=12,15,20$) reveal that the distribution evolves with time close to the transition, $w=12$, but is stationary for stronger disorder. Furthermore, the exponent $\zeta$ decreases as $w$ increases. At very strong disorder, $\zeta$ tends to $\zeta_\infty =1$. This is in stark contrast to the localization transition on Random Regular Graphs\,\cite{DeLucaKravtsov2014}, where $\zeta = 1/2$ at criticality and decreases further upon entering the localized phase. A real-space picture of $\zeta_\infty =1$ involves independently flipping pairs of nearest neighbor spins; see Sec.~\ref{subsec:exponent_zeta} 
of SI for details, or Ref. \cite{Leticia2024} for a possibly related process on Erdős-Rényi graphs. The results for the distribution function are limited to times $Jt_{max}=20\pi$ due to finite value of $T_1$ in our system. We have checked numerically that one does not observe significant time evolution at longer times, see Sec.~\ref{SI_finite_times} of the SI for details.

% \noindent\textcolor{blue}{VII.} This result can be understood by noting that at low $w$, the configurations are uniformly distributed. As $w$ increases, the number of actual bit-flips decreases while the probability distribution becomes increasingly peaked, as we will show below. For non-ergodic states, this sharpening suggests the formation of distinct spatial structures at high disorder, in agreement with Ref.~\cite{Leticia2024}.  

Thus far we have described the dynamics of wavefunctions without appealing to their underlying structure in configuration space. Next, we provide a new perspective on the data in Figs. \ref{R(t)} and \ref{WavefuncStats}---by constructing graphs capturing the effective Hilbert space connectivity, with the goal of visualizing wavefunction spreading and understanding how disorder `prunes' the graph. In Fig.~\ref{Network}a, we show the highest-probability bitstrings (rank 1), along with several other less likely bitstrings. These bitstrings are ordered in Fig.~\ref{Network}b according to their Hamming distance with respect to the most likely bitstring. 
Representing the data as a Wave-Function Network (WFN) \cite{Network_PRX2024,andreoni2025network} offers an insight on the geometrical structure of the set of bitstrings, and allows us to infer the distribution over the average degree of connectivity of the graph. In Fig. \ref{Network}c we embed the graphs at both low ($w=8$) and high ($w=20$) disorder in principal component space. A qualitative difference can already be observed,
which is quantitatively captured by the degree distribution $\mathcal{P}(k)$, shown in Fig. \ref{Network}d. At small disorder, the graph connectivity is compatible with that of a random geometric network. Conversely, the connectivity structure of graphs considered at large disorder is governed by self-connections, and shows a scale-free behavior over at least two orders of magnitude, as expected from the probability distribution of bitstrings. 

We now connect the wavefunction statistics collected in Fig.\,\ref{WavefuncStats} to this configuration space graph description. Using the Hamming distance between two bitstrings, $r = d(z',z)$, we define $N_{\rm shell}(r)$ as the number of bitstring pairs separated by a distance $r$. As shown in Fig. \ref{Graph_analysis}a, we empirically find that $N_\text{shell} (r)$ peaks around a value $r^*$ that shifts with increasing disorder away from the fully equiprobable ergodic limit of $r = n/2$. Another empirical observation is that the conditional probabilities $\mathcal{P}(x|r)\sim x^{-\alpha_r-1}$ are each broadly distributed (see Fig.\ref{Graph_analysis}b and SI), where we again define $x\equiv N|\psi|^2$. Combining with the expression above, the log-bin probability distribution is:
\begin{equation}
\mathcal{P}(x) \propto \sum_r N_\text{shell}(r) x^{-\alpha_r-1},
\end{equation}
\noindent Because $N_\text{shell}(r)$ is sharply peaked, it turns out that we can approximate the full distribution $\mathcal{P}(x)$ by the value of its marginal distribution at $r_*$ (see SI). This means that $\zeta\approx \alpha_{r_*}$, to high accuracy. For $\zeta_\infty<1$, the wavefunction distribution in the shell at $r_*$ is biased towards a finite number of bitstrings with large probability. In this regime, even though there are many low-probability bitstrings, their number does not grow fast enough to compensate their smaller probability weight; this situation corresponds to the localized limit. One therefore finds that $\zeta_\infty=1$ is the marginal exponent separating ``few dominant configurations'' from ``an extensive number of configurations, each contributing with small probability" (see SI for more details). This process is schematized in the right panel of Fig.\,\ref{Graph_analysis}c.

We can further characterize the scaling properties of many-body wave functions near the localization and non-ergodic transitions with a multifractal analysis. 
% Here, we apply the recently-developed Binary Intrinsic Dimension (BID) estimator}\cite{acevedo2025unsupervised}  to provide a volumetric view of ergodicity (or lack thereof) directly within configuration space. The BID explicitly captures how space-filling the wave function is. 
Here, to provide a volumetric view of ergodicity (or lack thereof) directly within configuration space, we exploit the concept of intrinsic dimension, a proxy of algorithmic complexity that quantifies the minimum number of independent variables needed to describe the set of collected configurations. To compute this quantity, we apply the recently-developed Binary Intrinsic Dimension (BID) estimator\cite{acevedo2025unsupervised}. BID explicitly captures how space-filling the wave function is, by inferring the effective dimensionality of a set of data-points living on the edges of an hypercube from the distribution of Hamming distances. 
For an ergodic phase, one expects that the BID, normalized by system size, saturates to 1 as the volume is increased---thereby recovering the full Hilbert space. Conversely, the normalized BID of a non-ergodic state will saturate to a finite fraction less than 1, suggesting that the dynamics are restricted to a subset of the full Hilbert space. In Fig. \ref{Graph_analysis}d, we present a finite-size scaling analysis of the BID. These data distinguish between two physically relevant regimes: (1) at low disorder, an ergodic phase in which the effective dimension saturates to the full embedding dimension ($d_0/n \rightarrow 1$); and (2) a stable multifractal regime at intermediate disorder where the scaling dimension remains robustly smaller than 1. At the largest disorder strength considered, the BID is very weakly dependent on system size, suggesting proximity to localization~\cite{erazo2026dimensionalityhopfieldmodel}
. Further support for multifractality is provided by Fig. \ref{Network}d; the scale-free distribution of the effective graph strongly suggests wave-function self-similarity at all scales. Analogous behavior has been observed for conformal field theories as well\,\cite{Network_PRX2024,andreoni2025network}.

To date, studies of ergodicity breaking and many-body localization have predominantly focused on one dimension, whereas in higher dimensions the stability of non-ergodic phases remains contentious~\cite{pal2010many,Krav2015,Krav2019,tikhonov2019statistics,Biroli2021,lunkin2025local, Safonova2,QREM,Smelyan2020, Galitski2022,BarLev2015, Agarwal2015,Alet2019,Pino2016, Pino2017,Laumann2017,gopalakrishnan2019instability,suntajs2020quantum,Long2023,Yin_2024,Tarzia2020,BiroliTarzia2024,placke2024topological,Hong_2025,de2024absence, Doggen2020, LiWahl2024, LiWahl2025,vanoni2024renormalization, altshuler2025renormalization}. By resolving configuration-basis dynamics, we demonstrate the absence of a direct transition to a localized phase, revealing instead an extended non-ergodic regime beginning near $w \sim 10$ across all experimentally accessible system sizes and timescales. Within this regime, we observe reinforcing signatures of ergodicity breaking, including Hilbert-space network reorganization~\cite{Network_PRX2024,andreoni2025network}, multifractality, and algebraic relaxation. Reconstructing effective configuration graphs directly from processor snapshots establishes a promising framework to probe high-dimensional quantum states, demonstrating that quantum hardware can isolate microscopic mechanisms governing non-equilibrium phases inaccessible to classical numerics.

\vspace{3mm}
\noindent\textbf{Acknowledgment.} We are grateful to D. Bhakuni, M. Hopjan, V. Kravtsov, D. Long, V. Khemani, and L. Vidmar for useful discussions.

\newpage
\onecolumngrid
\vspace{5mm}

% \newpage
% \onecolumngrid

% \vspace{1em}
\begin{flushleft}

{\hypertarget{authorlist}{${}^\dagger$}  \small Google Quantum AI and Collaborators}

\bigskip

    \renewcommand{\author}[2]{#1\textsuperscript{\textrm{\scriptsize #2}}}
    \renewcommand{\affiliation}[2]{\textsuperscript{\textrm{\scriptsize #1} #2} \\}
    \newcommand{\corrauthora}[2]{#1$^{\textrm{\scriptsize #2}, \hyperlink{corra}{\ddagger}}$}
    \newcommand{\corrauthorb}[2]{#1$^{\textrm{\scriptsize #2}, \hyperlink{corrb}{\mathsection}}$}

\begin{footnotesize}

\newcommand{\xCENN}{\affiliation{1}{Nanocenter CENN, Ljubljana, Slovenia}}

\newcommand{\xGoogle}{\affiliation{2}{Google Research, Mountain View, CA, USA}}

\newcommand{\xStanford}{\affiliation{3}{Department of Applied Physics, Stanford University, Stanford, California 94305, USA}}

\newcommand{\xSISSA}{\affiliation{4}{SISSA,International School for Advanced Studies, via Bonomea 265, 34136 Trieste, Italy}}

\newcommand{\xICTP}{\affiliation{5}{ICTP, The Abdus Salam International Centre for Theoretical Physics, Strada Costiera 11, 34151 Trieste, Italy}}

\newcommand{\xPrinceton}{\affiliation{6}{Department of Electrical and Computer Engineering,
Princeton University, Princeton, NJ, USA}}

\newcommand{\xStanfordPhys}{\affiliation{7}{Department of Physics, Stanford University, Stanford California 94305, USA}}

\newcommand{\xCornell}{\affiliation{8}{School of Applied and Engineering Physics, Cornell University, Ithaca, New York 14853, USA}}

\newcommand{\xOxford}{\affiliation{9}{Department of Physics, Clarendon Laboratory, University of Oxford, OX1 3PU, UK}}

\newcommand{\xUConnStorrs}{\affiliation{10}{Department of Physics, University of Connecticut, Storrs, CT}}

\newcommand{\xUMass}{\affiliation{11}{Department of Electrical and Computer Engineering, University of Massachusetts, Amherst, MA}}

\newcommand{\xUCSB}{\affiliation{12}{Department of Physics, University of California, Santa Barbara, CA}}

\newcommand{\xUCRECE}{\affiliation{13}{Department of Electrical and Computer Engineering, University of California, Riverside, CA}}

\newcommand{\xPritzker}{\affiliation{14}{Pritzker School of Molecular Engineering, University of Chicago, Chicago, IL}}

\newcommand{\xUCRPA}{\affiliation{15}{Department of Physics and Astronomy, University of California, Riverside, CA}}

\newcommand{\xAuburnECE}{\affiliation{16}{Department of Electrical and Computer Engineering, Auburn University, Auburn, AL}}

\newcommand{\xJSI}{\affiliation{17}{Jo\v{z}ef Stefan Institute, Ljubljana, Slovenia}}

\newcommand{\xUniBO}{\affiliation{18}{Dipartimento di Fisica e Astronomia, Università di Bologna, via Irnerio 46, I-40126 Bologna, Italy}}

\newcommand{\xUAug}{\affiliation{19}{Theoretical Physics III, Center for Electronic Correlations and Magnetism, Institute of Physics, University of Augsburg, 86135 Augsburg, Germany}}

\newcommand{\xCAAPS}{\affiliation{20}{Centre for Advanced Analytics and Predictive Sciences (CAAPS), University of Augsburg, Universitätsstr.\ 12a, 86159 Augsburg, Germany}}

% Make sure the numbers match up ^^ vv

\newcommand{\CENN}{1}
\newcommand{\Google}{2}
\newcommand{\Stanford}{3}
\newcommand{\SISSA}{4}
\newcommand{\ICTP}{5}
\newcommand{\Princeton}{6}
\newcommand{\StanfordPhys}{7}
\newcommand{\Cornell}{8}
\newcommand{\Oxford}{9}
\newcommand{\UConnStorrs}{10}
\newcommand{\UMass}{11}
\newcommand{\UCSB}{12}
\newcommand{\UCRECE}{13}
\newcommand{\Pritzker}{14}
\newcommand{\UCRPA}{15}
\newcommand{\AuburnECE}{16}
\newcommand{\JSI}{17}
\newcommand{\UniBO}{18}
\newcommand{\UAug}{19}
\newcommand{\CAAPS}{20}

\corrauthora{A. Lunkin}{\CENN},
\corrauthora{N. S. ~Ticea}{\Google, \!\Stanford},
\corrauthora{M. T. ~Solomon}{\Google, \!\Pritzker},
\corrauthora{R. Andreoni}{\SISSA, \!\ICTP},
\author{S. ~Kumar}{\Google, \!\Princeton},
\author{C. ~Miao}{\Google, \!\StanfordPhys},
\author{J. ~Choi}{\Google, \!\Cornell},
\author{M. Alghadeer}{\Google, \!\Oxford},
\author{I. Drozdov}{\Google,\! \UConnStorrs},
\author{D. Abanin}{\Google},
\author{A. Abbas}{\Google},
\author{R. Acharya}{\Google},
\author{L. Aghababaie~Beni}{\Google},
\author{G. Aigeldinger}{\Google},
% ...
\corrauthorb{P. Roushan}{\Google},
\corrauthorb{M. V. Feigel'man}{\CENN, \!\JSI},  
\corrauthorb{L. B.~Ioffe}{\Google},

\bigskip

\xCENN
\xGoogle
\xStanford
\xSISSA
\xICTP
\xPrinceton
\xStanfordPhys
\xCornell
\xOxford
\xUConnStorrs
\xUMass
\xUCSB
\xUCRECE
\xPritzker
\xUCRPA
\xAuburnECE
\xJSI
\xUniBO
\xUAug
\xCAAPS

\vspace{2mm}

{\hypertarget{corra}{${}^\ddagger$} These authors contributed equally to this work.}\\

{\hypertarget{corrb}{${}^\mathsection$} Corresponding authors: pedramr@google.com, Mikhail.Feigelman@nanocenter.si, and ioffel@google.com}\\

\end{footnotesize}
\end{flushleft}

\twocolumngrid

\bibliography{References}

\makeatother

\renewcommand{\abstractname}{\vspace{+\baselineskip}}
\makeatletter
\renewcommand{\thesection}{\arabic{section}}
\renewcommand{\thesubsection}{\thesection.\Alph{subsection}}
\renewcommand{\thesubsubsection}{\alph{subsubsection}}
\renewcommand{\thefigure}{S\@arabic\c@figure}
\renewcommand{\theequation}{S\@arabic\c@equation}
\renewcommand{\thetable}{S\@arabic\c@table}

\makeatletter
\renewcommand{\p@subsection}{}
\makeatother

\onecolumngrid

\title{Supplementary Materials for\\``Hilbert-space signatures of non-ergodic glassy dynamics''}
\author{Google Quantum AI and Collaborators}
% \date{\today}

\maketitle

\tableofcontents

\newpage

\section{List of symbols}
\label{sec:symbols}
\renewcommand{\arraystretch}{1.5}
\begin{center}
\begin{tabular}{ |l|l| } 
 \hline
 Symbol & Description\\ 
 \hline
$J$ & Spin-spin coupling strength (XY interaction) \\
$H$ & Hamiltonian of the spin-$\frac{1}{2}$ model on the Cayley tree \\
$\sigma^a_i$ & Pauli matrices (where $a=x,y,z$ or $1,2,3$), $\sigma^a_i = 2S^a_i$ \\
$\sigma^\pm_i$ & Raising and lowering spin operators, $S^\pm_i$ \\
$h_i$ & Random magnetic field at site $i$ \\
$W$ & Full width of the box distribution for $h_i$ \\
$n$ & Total number of spins (or system size) \\
$K$ & Branching number of the Cayley tree (CT) \\
$Z$ & Full coordination number of sites on CT, $Z=K+1$ \\
$\mathcal{H}$ & Operator $\hat{1} \otimes H^T - H \otimes \hat{1}$ in the extended Hilbert space \\
$|\hat{O}\rangle$ & Operator $\hat{O}$ represented as a wave function in the extended space \\
$\mathcal{V}$ & Perturbation operator in the extended space, related to $V$ \\
$\Gamma_i^r, \Gamma_j^{\pm}$ & Local relaxation rate for spin at site $i$ (or $j$) \\
$\Sigma_z(\epsilon)$ & Self-energy for the $z$-component of spin operator \\
$\Sigma_\pm(\epsilon)$ & Self-energy for the $\sigma^\pm$ components \\
$U$ & Anti-hermitian operator for Schrieffer - Wolff transformation \\
$H_{eff}$ & Effective Hamiltonian after Schrieffer - Wolff transformation \\
$V_{eff}$ & Effective perturbation Hamiltonian, $\frac{1}{2}[V,U]$ \\
$\Gamma_j^\phi$ & Local dephasing rate for spin at site $j$ \\
$J_r, w_r$ & Critical coupling strength and dimensionless disorder ($w_r=W/J_r$) for relaxation \\
$J_\phi, w_\phi$ & Critical coupling strength and dimensionless disorder ($w_\phi=W/J_\phi$) for dephasing \\
$F(x)$ & Function used to determine the critical point of linear recursions \\
$P_0(h)$ & Bare distribution function for the local field $h$ \\
$P_1(h)$ & Effective distribution function for $h$ accounting for self-energy correction (relaxation) \\
$P_2(h_1,h_2)$ & Renormalized distribution function for $h_1, h_2$ (dephasing) \\
$G(e^{-x})$ & Laplace transform of the probability density function $\mathcal{P}(\Gamma_j)$ \\
$S$ & $\Re\Sigma$, the real part of the self-energy correction (dephasing) \\
\hline
\end{tabular}
\end{center}

\newpage

\section{Experimental details and extended data}\label{sec:exp_tech}
\subsection{Procedure}
The experiments are performed on a Willow device architecture as presented in Ref.~\cite{google2025quantum}, using the calibration protocols shown in Ref.~\cite{andersen2025thermalization} to set the coupling rates and $Z$-fields of the analog Hamiltonian. Errors in the $XY$-couplings and the $Z$-fields are typically well below 1\% of the disorder strengths studied. Moreover, using a coupling strength of $g=5$ MHz, higher-order Hamiltonian terms (such as $ZZ$-interactions) are on the order of 100s of kHz and do therefore not impact our conclusions. 

In all experiments, the qubit frequencies are first ramped quickly ($\sim 1$ ns) to the values corresponding to the relevant disorder instance, before ramping up the coupling rates over another $\sim 1$ ns. The system is then evolved for a variable amount of time under the analog Hamiltonian, followed by a 1 ns Hamiltonian off-ramp, and finally measurements in the $z$-basis. Since the Hamiltonian is photon number conserving, we mitigate the effects of photon loss by postselecting the data, removing any bitstrings in which the photon number is not conserved. This is the primary reason we do not probe times longer than $Jt=100$, since the postselection ratio becomes prohibitively low after this time. While postselection mitigates T1 errors, we note that dephasing errors are not removed by this technique; nevertheless, we do not observe substantial effects of dephasing on the bitstring probabilities in Fig. 3 in the main text, which would otherwise be driven towards uniform distributions. Moreover, if dephasing played a dominant role, one would expect to find $\eta\sim n$ in Fig. 2d, in stark contrast to the exponent of 2.4 observed in our experiment.

\subsection{Glassy behaviour in magnetization.}

\begin{figure*}[t!]
\includegraphics[width=\textwidth]{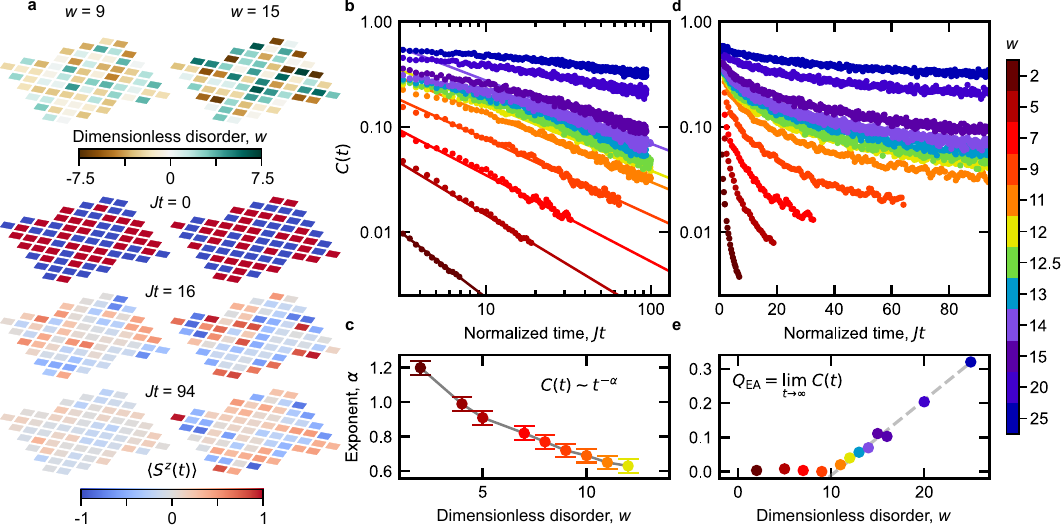}
\centering
\caption{\textbf{Evidence of glass formation in the frozen magnetization. a,} Temporal evolution of magnetization for disorders of $w=9$ and $w=15$ (corresponding disorder patterns showed on top), revealing slower relaxation at the higher disorder. \textbf{b,} Time dependence of the spin glass correlation function $C(t)$, showing power law relaxation of $C(t)$ in the ergodic phase.  
\textbf{c,} Disorder-dependence of the dynamic exponent $\alpha$ that describes the power-law decay of $C(t)$. \textbf{d,} Same as \textbf{b}, but with linear axis, revealing the absence of complete relaxation at higher disorders. \textbf{e,} The Edwards-Anderson order parameter, $Q_\text{EA} = C(\infty)$, as a function of $w$ shows an abrupt change in slope near $w=w_c\simeq 10$. $Q_\text{EA}$ is extracted from fitting $C(t)$ with a decaying functional form (exponential with or without a power-law prefactor, see SI) with a constant offset.} \label{fig:GlassFormation}
\end{figure*}

We accompany our experimental studies by measuring the local magnetization in real space, a quantity that is readily accessible on most quantum processors. 
Starting from a 59-qubit product state at $t=0$, we consider the relaxation of the spin distributions, shown in Fig.~\ref{fig:GlassFormation}a for two different disorder strengths, $w=9$ and $w=15$. The subsequent spatial maps show that the magnetization profile relaxes almost completely in the case of low disorder ($w=9$), whereas substantial magnetization remains frozen at higher disorder ($w=15$). This is a first indication of reduced ergodicity. To further study this contrasting behaviour, we next plot the dynamic spin-glass correlation function, 
\begin{equation}
C(t) = \overline{\langle S_i^z(0) S_i^z(t)\rangle^2}, 
\end{equation}
for a wide range of disorder values (Fig.~\ref{fig:GlassFormation}b). In the weakly disordered regime, the correlation function is found to obey a power-law decay $C_{fit}(t)=a t^{-\alpha}$, with an exponent $\alpha$ that decreases from unity to about $ 2/3$ as $w$ approaches a critical value, $w_{\mathrm{c}}\sim 10$ (Fig.~\ref{fig:GlassFormation}c). For values above $w_c$, however, the decay deviates from power-law behavior and saturates at a non-zero value of the Edwards-Anderson order parameter, $Q_{\mathrm{EA}} = C(t\to\infty)$, as is more clearly observed with a linear time axis (Fig.~\ref{fig:GlassFormation}d). This dependence  fits well exponential law at with non-zero offset: 
\begin{equation}C_{fit}(t)=a \exp(-\Gamma' t) + Q_{EA}\,\qquad  \text{or} \qquad C_{fit}(t)=a t^{-\alpha} \exp(-\Gamma' t) + Q_{EA}
\end{equation}
close to the transition. This behaviour is quantified in Fig.~\ref{fig:GlassFormation}e, where we show the extracted $Q_{\mathrm{EA}}$: while $Q_{\mathrm{EA}}$ is found to be very small in the ergodic regime, we observe a clear change to linear growth with disorder for $w \gtrsim w_c$, consistent with a transition or sharp cross-over. The observed freezing-in of magnetization dynamics is in qualitative agreement with earlier observations~\cite{choi2016}.

\begin{figure}[h!]
    \centering
    \includegraphics{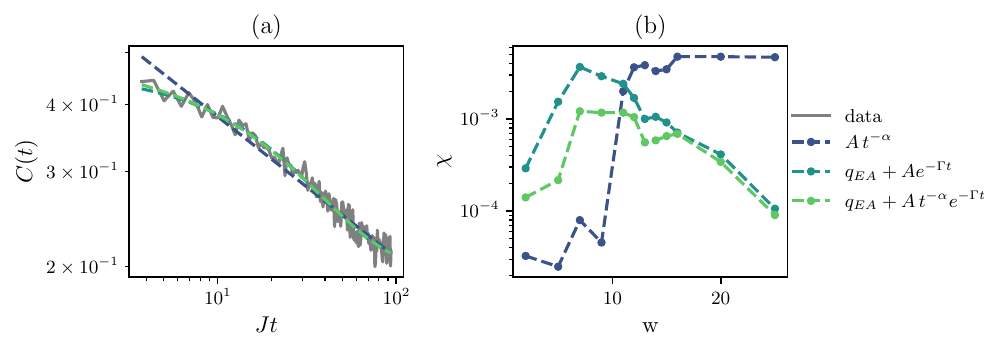}
    \caption{Comparison of remanent magnetization fits by different time dependencies. Left panel: data for $n=59$ $w=14$ and their fits, with and without offset. The data shown here were filtered with a degree $r=2$  Savitsky Golay filter. Right panel: data deviation from different fits as a function of $w$ for $n=59$. For lower disorder, a power-law fit is preferred, whereas at higher disorder an exponential fit with a constant offset provides a better description. In this regime, it is not possible to distinguish between a purely exponential fit and one that includes a power-law prefactor. Note that, to distinguish this change in behavior, we assume that $q_{EA}$ should be greater than 0.01. For this reason, the power-law model without an offset is preferred at lower disorder.}
    \label{fig:fit_errors}
\end{figure}

The quality of the fit can be quantified by comparison of  the actual data with the fits and define quality of fit, $\chi$ by
\begin{equation}
\chi=\frac{1}{n_t} \sum_t \left( \frac{C_{fit}(t)-C_t}{C_t} \right)^2
\end{equation}
where $n_t$ is the number of data points in time, $C_{fit}(t)$ is the fitting function and $C_t$ is filtered value of measured magnetization squared after time $t$, $m^2_t = (1/n) \sum_i \langle \sigma_i^z(t) \rangle^2 $. The measured values of $m^2_t$ fluctuates significantly over time, even after being averaged over ~100 realizations. These fluctuations dominate the fit quality $\chi$ at large disorders; to suppress this effect, we smoothened the data by applying the Savitsky Golay filter, $G_r$, with degree $r=2-5$ to the data: $C_t=\sum_t' G_{r}(t-t') m^2_t$. We emphasize that the fit parameters are not affected by the application of the filter; it affects only the evaluation of the fit quality, $\chi$. Slightly above the transition the best fit is achieved by an exponential with a power law prefactor and constant offset, see Fig.\ref{fig:fit_errors}a. Below the transition, at $w<w_c\sim10$, the power law fit is preferred, see Fig.\ref{fig:fit_errors}b. Similarly, exponential fits with constant offset are poor below the transition. Far above the transition, exponential and exponential with power law prefactor fits perform equally well. In this regime, the exponent $\alpha$ of the power law becomes very small, e.g. $\alpha=0.15$ for $w=20$, which makes it very difficult to distinguish the fits. The values of the constant offset for these two fits are practically identical above the transition. 

\vspace{30mm}

% \newpage
\subsection{Breakdown of diffusion.}
\begin{figure*}[h!]
\includegraphics[width=\textwidth]{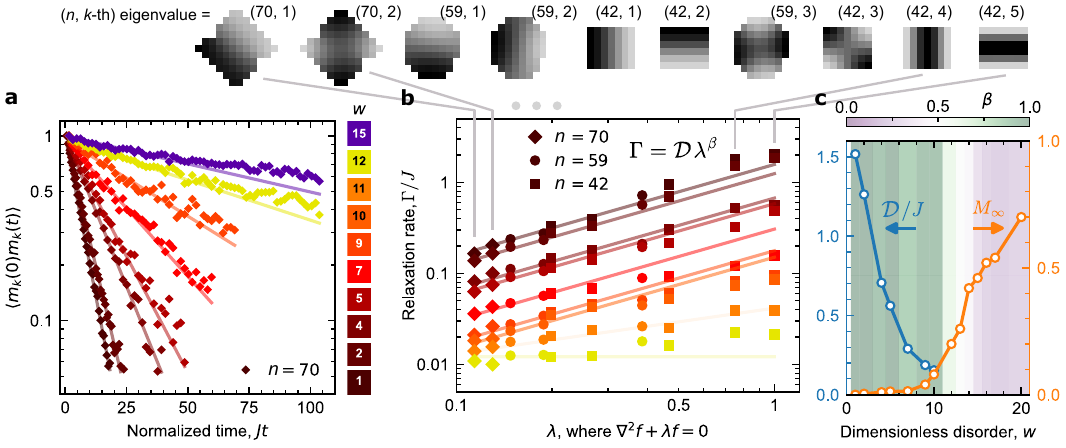}
\centering
\caption{\textbf{Disappearance of diffusion at strong disorder. a,}
 The measured relaxation of magnetization $\langle m_k(t)m_k(0)\rangle$ for the smallest eigenmode ($k=1$) at several values of disorder ($n=70$). \textbf{b,} 
 Extracted relaxation rates for $n=42$, $59$ and $70$. Solid lines are fits using $\Gamma = \mathcal{D} \lambda$. \textbf{c,} The extracted diffusion constant $\mathcal{D}$ (left axis) and non-zero frozen amplitude $M_\infty$ (right axis). Background colours denote $\beta$, resulting from fitting data in \textbf{b} with  $\Gamma = \mathcal{D} \lambda^\beta$. Diffusion applies for $\beta \approx 1$ and fails as $\beta$ departs from unity and hence $\mathcal{D}$ cannot be extracted above $w>10$.} \label{Fig4}
\end{figure*}

 In the presence of locally conserved quantities, the relaxation of an ergodic state can be characterized by how each conserved quantity diffuses. Figure \ref{Fig4}a shows the relaxation of the normalized magnetization correlations, $C_1$ where $C_k=(1/R_{k})\langle m_{k}(t)m_{k}(0)\rangle$, $R_k=\langle m_{k}^2(0)\rangle$, for various disorder strengths, after projecting (in post-processing) onto eigenmodes of the diffusion operator $-\nabla^2$ (the modes are indexed by $k$ and take on eigenvalues $\lambda$; owing to the irregular lattice geometry, they are computed numerically for each system size considered here, e.g., $\lambda_{\mathrm{min}}=0.114$ for $n=70$). At weak disorder ($w=1,2,4$), the magnetization decays exponentially $\propto e^{-\Gamma t}$. For moderate disorder ($w=5,7,9$), the relaxation becomes slower, pointing to the onset of anomalous diffusion. At strong disorder, the decay is slower than exponential at long times, and the magnetization does not fully vanish; even at the longest time scales, a finite remanent amplitude $M_\infty$ persists. An ``effective'' relaxation rate $\Gamma$ can still be defined from the short-time slope of the relaxation curve. Fig.\,\ref{Fig4}b shows the fitted $\Gamma$ versus the Laplacian eigenvalue, $\lambda$, 
for several disorder strengths and three different system sizes of 42, 59 and 70 qubits ($50-100$ disorder realizations for each case), which show good agreement for similar values of $\lambda$.  At low disorder, $\Gamma$ scales linearly with the eigenvalue up to $\lambda \approx 0.5$, consistent with normal diffusion with diffusion constant $\mathcal{D}$. At stronger disorder, this proportionality weakens and eventually breaks down. Fig.\,\ref{Fig4}c summarizes the extracted diffusion constant $\mathcal{D}$ versus disorder strength, showing that near a critical value $ w\simeq 10$, $\mathcal{D}$ drops to $\mathcal{D} \sim 0.1$, while a finite residual magnetization $M_\infty$ emerges for $w>w_c$. Here we defined that residual magnetization as the infinite time limit of the eigenmode $k$ for normalized correlation function, $C_k$,  $M_\infty=C_k(t\rightarrow \infty)$. In this panel we also show the value of $\beta$ resulting from fitting $\Gamma = a \lambda^\beta$; as expected, $\beta\approx 1 $ in the ergodic phase but drops for $w>w_c$. 

\newpage
\section{Effect of finite time on wave function properties}
\label{SI_finite_times}
Due to finite values of the relaxation times, $T_1$ and $T_2$ the experimental data are limited to times $Jt<100$, with $Jt_{max}=20 \pi$ for largest system sizes (the effective $1/T_1$ for a many body system scales with the number of qubits in state $1$). Thus, it is important to check if our results on the fractal behavior of the wave function at $w>10$ correspond to transient time behavior or remain unchanged in the limit of infinite times. We checked numerically that for small system sizes that allow direct numerical computation the behavior of the wave function remains the same as time is increased.   
\begin{figure*}[h!]
\includegraphics[width=3.5in]{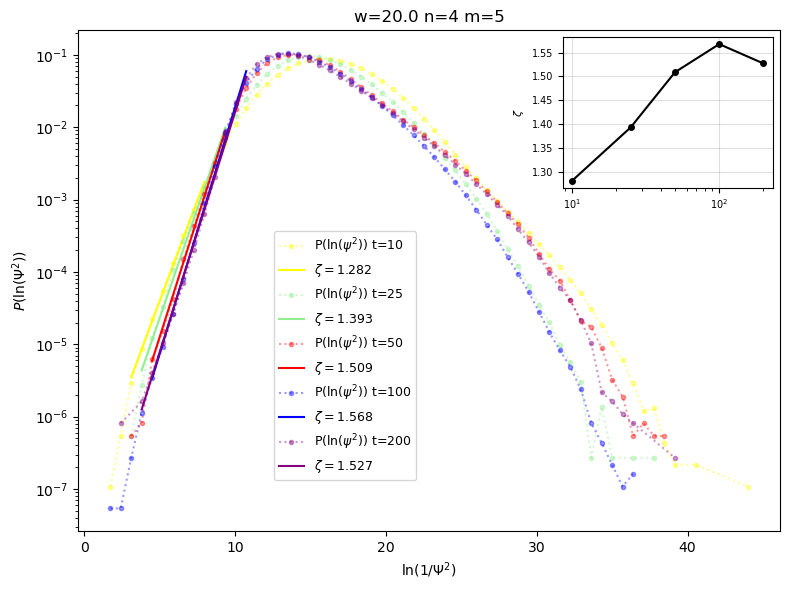}
\includegraphics[width=3.5in]{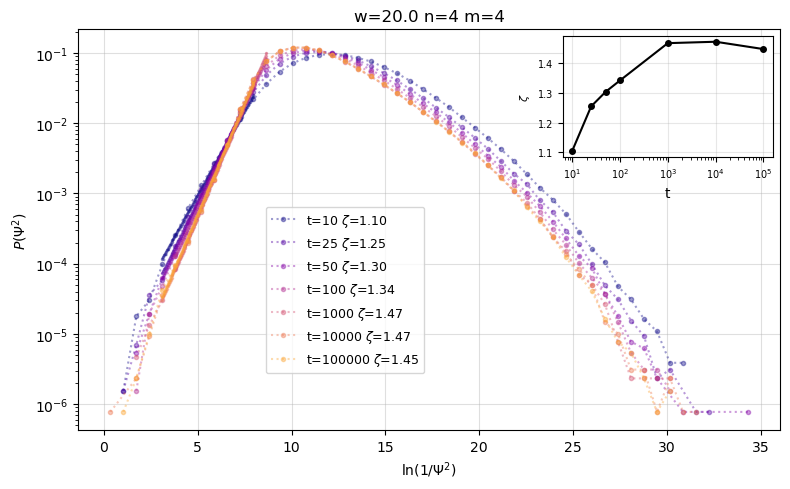}
\centering
\caption{\textbf{Distribution function} of $ln\Psi^2$ for $w=20$ for different times. The left panel shows the evolution of the wave function for 20 spin up to times $Jt=200$. The right panel shows similar evolution for 16 spin system up to times $t=10^5$} \label{FigPoPw20}
\end{figure*} 
We evaluated the finite time evolution using Chebyshev expansion for times up to $Jt=200$, see Fig.~\ref{FigPoPw20}, left panel, for 20 spin system at $w=20$. We observe very little evolution of the left tail of the distribution function above $Jt>50$ with exponent $\zeta$ that changes from $1.5$ at $Jt=50$ to $1.55$ at longest times. We have also checked for the presence of a very slow creep of the distribution function at times comparable to Heisenberg time, $t_H=1/\delta E$ where $\delta E$ is the distance between the levels, $\delta E \approx n J / 2^n$. We used full diagonalization of the Hamiltonian to evaluate time evolution for systems of 16 spins for which $t_H\sim 10^3$. We observe very little evolution of the left tail of the distribution function above $Jt=50$ with exponent $\zeta$ that varies between $1.3$ and $1.45$, see Fig.~\ref{FigPoPw20} right panel. Note that right tail of the distribution function, i.e. very small values of $\Psi^2$ shows much more significant evolution with time which should be expected. 
\begin{figure*}[h!]
\includegraphics[width=3.5in]{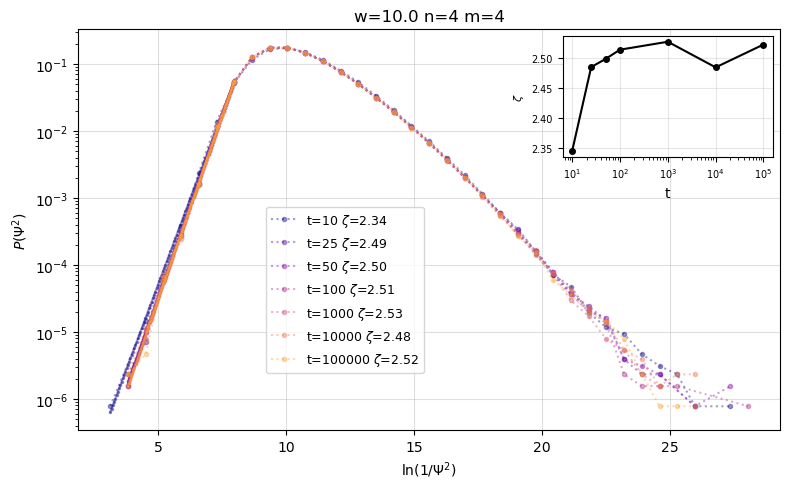}
\includegraphics[width=3.5in]{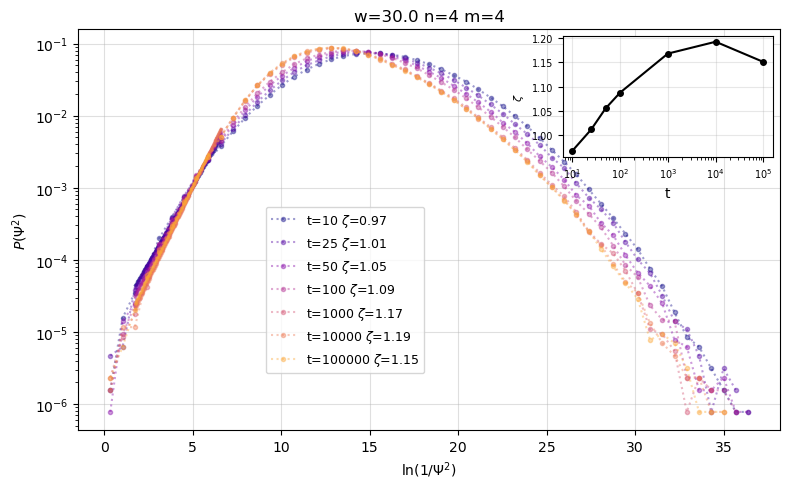}
\centering
\caption{\textbf{Distribution function} of $ln\Psi^2$ for $w=10$ (left panel) and $w=30$ (right panel) for different times for 16 spin system} \label{FigPoPn16}
\end{figure*} 

The value of $w=20$ corresponds to largest time dependence of the distribution function. For both smaller and larger $w$ the effect of time is smaller as one can see in Fig.~\ref{FigPoPn16}. For $w=10$ there is no appreciable variation of the exponent $\zeta$ that remains $\zeta=2.5\pm 0.02$ while for $w=30$ it changes between $\zeta=1.05$ and $\zeta=1.15$
We conclude that the qualitative behavior (power law dependence) of the distribution function remains the same at longest times while exponents obtained from the behavior at short times, $Jt~50$ might be 10\% less than the ones at longest times. 

% \newpage
\section{Extended data for the wavefunction network}

\begin{figure*}[th!]
\includegraphics[width=\textwidth]{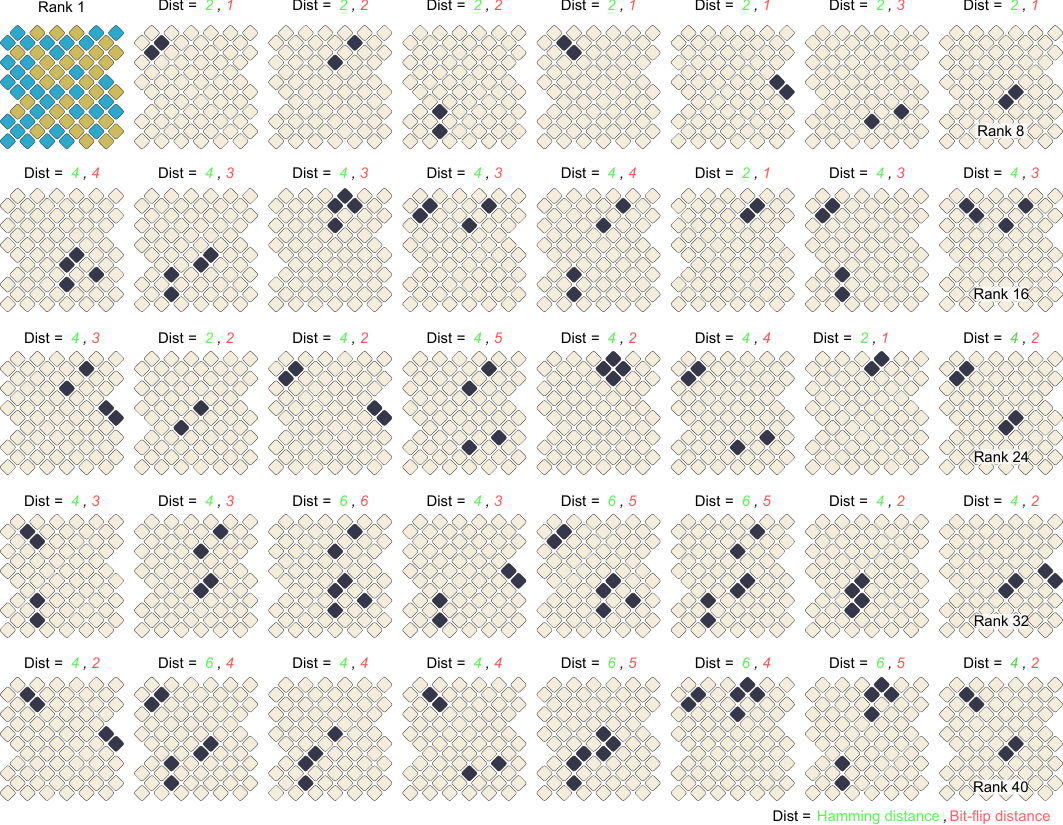}
\centering
\vspace{10mm}
\caption{\textbf{Difference maps of z-configs. }
The 40 most probable many-body configurations ($Z$-basis) at $Jt=60$ for a system of size $n=70$ at strong disorder ($w=35$). Configurations are ranked by their measured probability for a single disorder realization, with their structure shown as the bit-wise difference relative to the most likely state (Rank 1). From these difference maps, we compute two metrics relative to Rank 1: the Hamming distance (green), counting total mismatched bits, and the bit-flip distance (red), representing the minimum number of physical operations ($|i \rangle \leftrightarrow |j \rangle$) required to connect the states. These distances define the weights of the edges connecting nodes within the resulting wavefunction graph. The Hamming distance was used in computing clustering presented in the main text. }
\end{figure*}

\begin{figure*}[th!]
\includegraphics[width=0.8\textwidth]{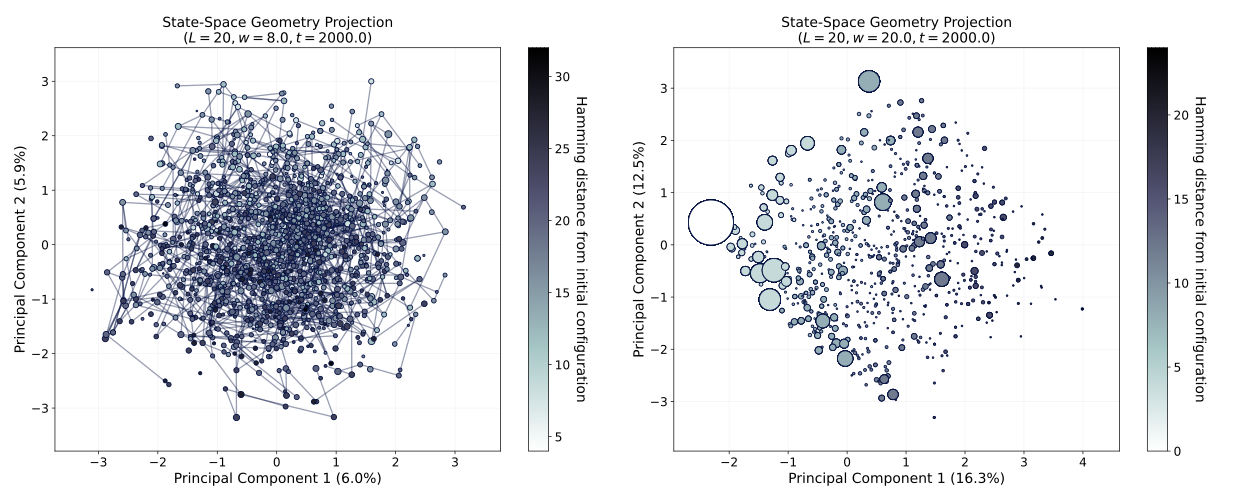}
\centering
\vspace{10mm}
\caption{\textbf{Wave-Function Networks. }Wave-Function Networks (WFNs)~\cite{Network_PRX2024, andreoni2025network} are a tool to study the structure of a set of configurations measured from a given wavefunction. They are constructed as described in the references: Nodes are identified by bitstrings embedded in the configuration space endowed with the Hamming distance; Links are drawn between pairs of nodes separated by less than a specific distance cutoff, which is chosen self-consistently as the average $n^{\mathrm{th}}$-nearest neighbor distance (here, we set $n=5$). The figure displays projections of a 1,000-node subsample from two WFNs onto their first two principal axes. These axes are derived via Principal Component Analysis (PCA) of the complete set of 10,000 $z$-bitstrings; the percentages associated with the $x$- and $y$-axes represent the fraction of the dataset's total variance explained by each principal component. Node sizes are proportional to their connectivities, and colors indicate the distance of each node from the initial bitstring. Both panels present representative data taken at $n=20$ and $Jt=60$, based on a single choice of disorder realization and initial bitstring. The left panel, corresponding to a low disorder of $w=8$, exhibits randomly distributed connections, a degree distribution with a short tail, and an absence of clear hubs. Graphs for all other considered disorder realizations and initial bitstrings share these analogous features. In the right panel, at a high disorder of $w=20$, the graph strongly clusters, allowing connections only between repeated bitstrings. Notice that nodes with a non-zero degree are actually overlapping nodes connected exclusively among themselves, representing these repeated bitstrings. This implies that node degrees are directly related to the repetition count of each bitstring in the dataset. Therefore, the information contained in the degree distribution $P(k)$ is identical to that contained in the probability distribution of the $z$-configurations. At high disorder, while many WFNs exhibit this similar clustered structure across different disorder realizations and initial bitstrings, we observe significantly more structural variability overall compared to the low-disorder case.}
\end{figure*}

\newpage
\section{Shell picture}
\begin{figure*}[]
\includegraphics[width=0.98\textwidth]{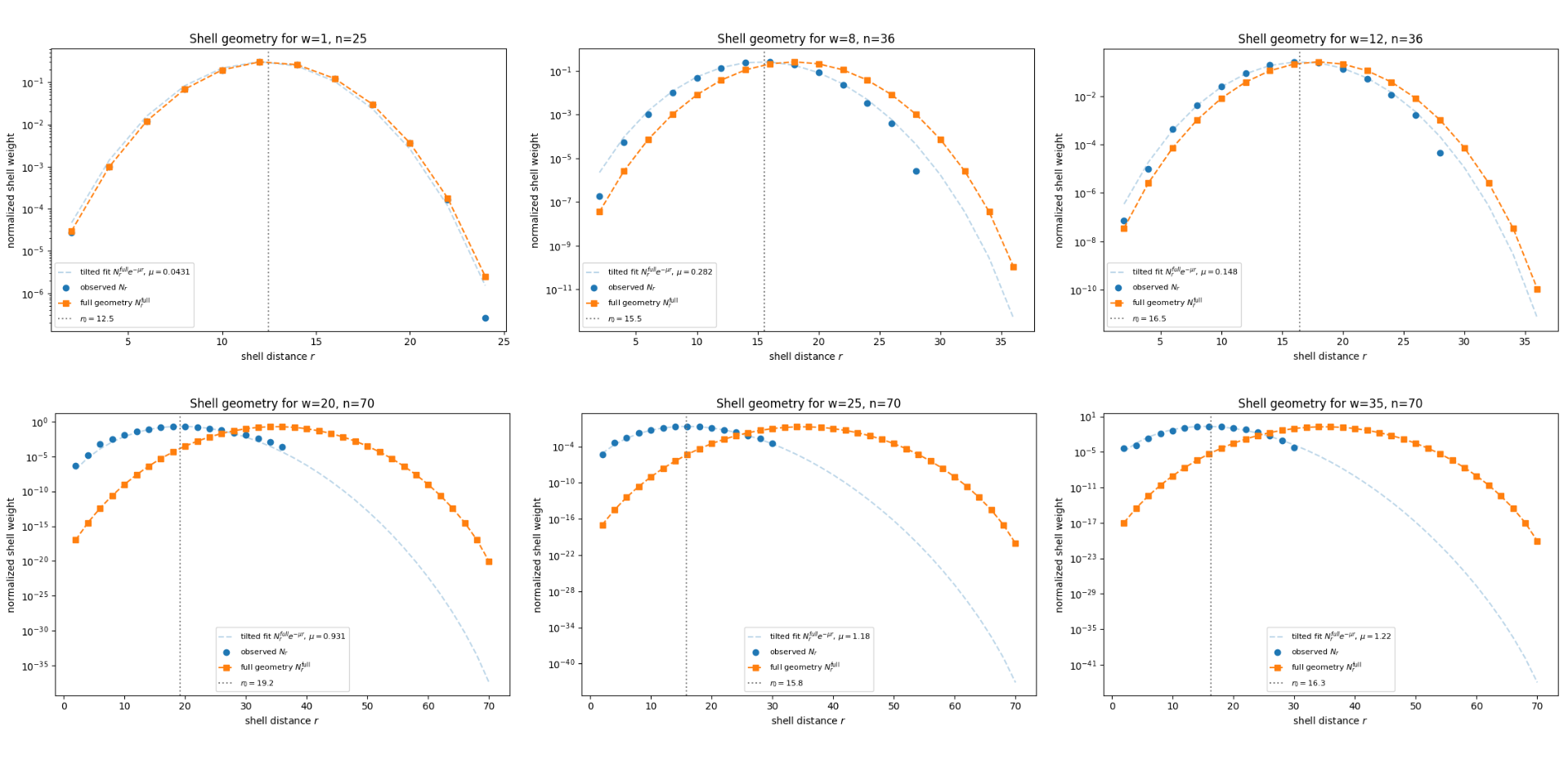}
\centering
\caption{\textbf{Shell geometry} For every single pair of bitstrings, we calculate the Hamming distance $r$ between them and then plot the histogram, normalized so that $\sum_r N_r=1$, of all pairwise distances as a function of $r$. We also show, as reference, the normalized distribution over $N_r$ assuming each bitstring in Hilbert space appears once (orange). Finally, we include a modified fit (pale blue) in which the full shell is tilted by an amount $\mu$ as described in Eq. \ref{eq:shell_tilt}.} \label{fig:shell}
\end{figure*}

\begin{figure*}[]
\includegraphics[width=0.98\textwidth]{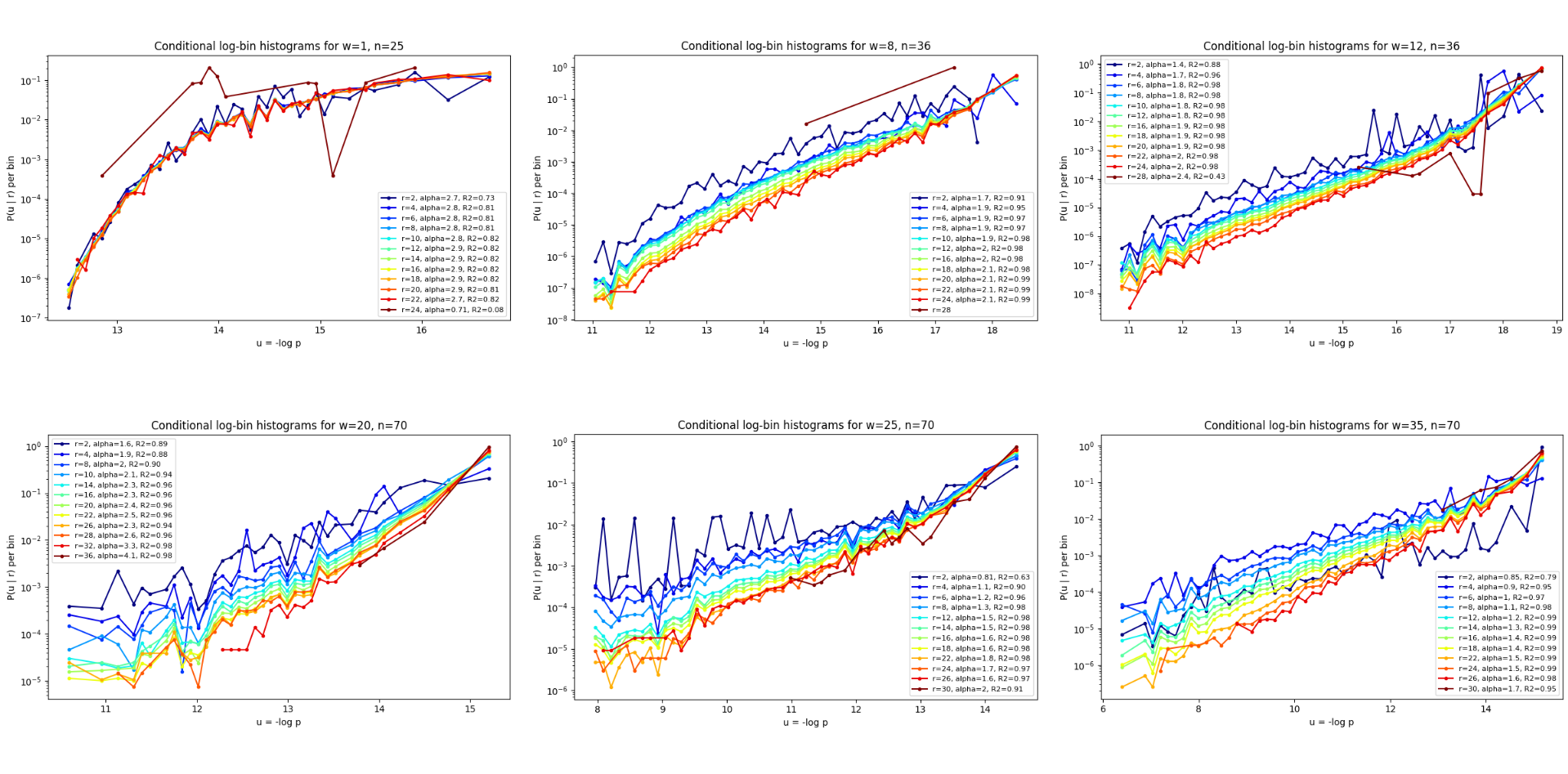}
\centering
\caption{\textbf{Conditional distribution} We plot the conditional distribution $P(u|r)$ as a function of $u$. Different colours correspond to distinct values of $r$. As the disorder strength increases, we see that the data follow an exponential form, particularly for intermediate values of $u$. To construct the histogram $H(r,u)$, we use the following formula: $H(r,u) = \sum_{ij} w_i w_j \bm{1}[d(z_i,z_j)=r]\bm{1}[u_j\in \text{bin centred at }u ]$, where $w_i$ and $w_j$ are the sampling/reweighting factors attached to bitstrings $i$ and $j$ (constructing the pairwise adjacency matrix between every pair of bitstrings in the sample is far too computationally intensive). In this way we do not bias ourselves with a reference bitstring and are able to collect more statistics. \label{fig:conditional_distribution}}
\end{figure*}

\begin{figure*}[]
\includegraphics[width=0.98\textwidth]{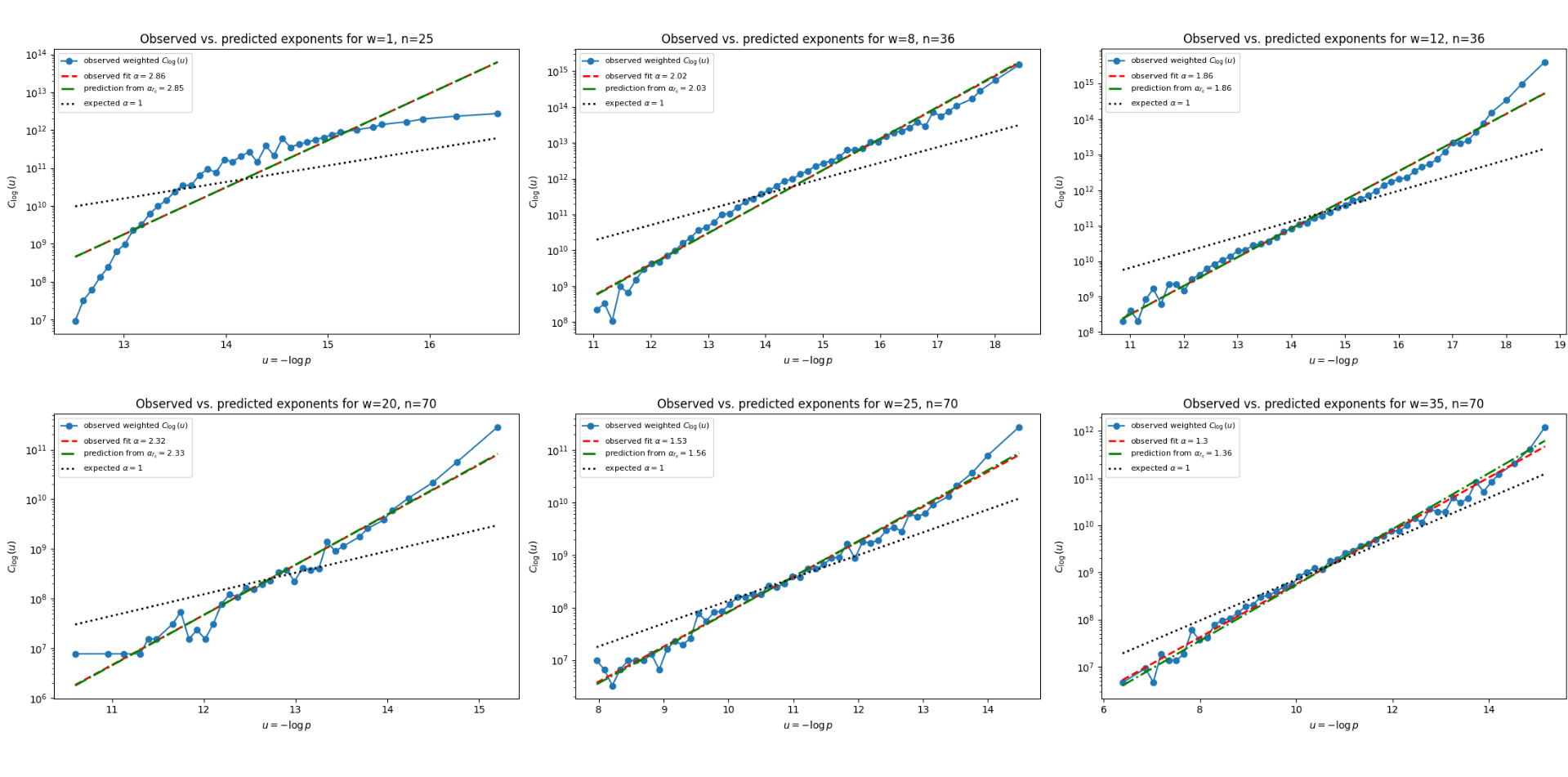}
\centering
\caption{\textbf{Predicted exponent} In this figure we plot the log-binned histogram (blue) alongside the best-fit exponent extracted from data (red) as well as the $\alpha_{r_*}$ prediction (green). Finally, for reference, we also show $\alpha=1$ (black dashed line), which corresponds to the limiting value $\zeta_\infty=1$. \label{fig:exponent}}
\end{figure*}

Allow us to define some metric for distance,
\begin{equation}
    r = d(z',z),
\end{equation}
which could, for example, measure the separation between two states in terms of Hamming distance (total number of flipped spins) or some notion of hopping distance (MWPM). We define $N_r^\text{shell}$ as the number of pairs of bitstrings that are separated by a distance $r$. Fix the metric to be Hamming distance. If we were to consider every single pair of bitstrings in Hilbert space, $N_r^\text{shell}$ would follow the distribution
\begin{equation}
    N_r^\text{shell} = \binom{L-M}{r/2}\binom{M}{r/2},
\end{equation}
where $M$ is the number of excitations \footnote{To see this, consider some fixed reference bitstring, say $111\cdots 10\cdots 000$. This bitstring has $M$ ones and $L-M$ zeros. To generate another bitstring at Hamming distance $r$, we must remove $r/2$ ones and fill $r/2$ zeros. There are $\binom{M}{r/2}$ ways of choosing which ones to void, and independently $\binom{L-M}{r/2}$ ways of picking which zeros to fill}. In our case, $M=L/2$, so
\begin{equation}
   N_r^\text{shell} = \binom{L/2}{r/2}^2
\end{equation}
For large $L$, we can approximate this quantity using Stirling's formula:
\begin{equation}
    N_r^\text{shell} \simeq \exp\left\{L H\left(\frac{r}{L}\right) \right\}; \qquad H(x) = -x\log x - (1-x)\log(1-x) \label{eq:shell_full}
\end{equation}
We recognize $H$ as the binary entropy function. The important thing to recognize is that $N_r^\text{shell}$ makes no assumptions about the underlying Hamiltonian or its dynamics. The only convention we had to fix was the metric. At this point we can make our first comparison to the dataset, plotting $N_r^\text{data}$ against $N_r^\text{shell}$. We present our results in Fig. \ref{fig:shell}. At low disorder, we find that $N_r^\text{data}\approx N_r^\text{shell}$. This means that the bitstrings we are sampling are representative of the full Hilbert space. But as the disorder strength increases, we find that the data distribution is increasingly skewed towards zero. It turns out that we can account for this drift by making a very simple modification to Eq. \ref{eq:shell_full}:
\begin{equation}
    N_r^\text{tilt}  \simeq \exp\left\{L H\left(\frac{r}{L}\right) -\mu r\right\}\label{eq:shell_tilt},
\end{equation}
where $\mu$ is some constant. These fits are overlaid with data.
To reconstruct the full distribution, we can sum over marginal distributions in $r$:
\begin{equation}
    P(u) = \sum_r N_r P(u|r)
\end{equation}
Experimentally, we observe the following distribution for $P(u|r)$, supported by Fig. \ref{fig:conditional_distribution}:
\begin{equation}
    P(u|r) = \frac{1}{Z_r}e^{\alpha_r u}\bm{1}_{[u_\text{min},u_\text{max}]},
\end{equation}
where $Z_r$ is a normalization constant and $\alpha_r$ is some exponent. It is interesting that not only is $P(u)$ distributed very broadly, but so is $P(u|r)$---that is to say, the distribution after conditioning on distance. Because $N_r$ is so sharply peaked around a single value of $r$, it turns out that the distribution can be very well approximated by
\begin{equation}
    P(u) \approx N_{r_*} P(u|r_*),
\end{equation}
where $r_*$ is the radial distance corresponding to the peak of $N_r$. This means that
\begin{equation}
    \zeta\approx \alpha_{r_*}
\end{equation}
We verify numerically that this approximation holds to very high accuracy in Fig. \ref{fig:exponent}.
\subsection{Probability mass balance}
We now comment on a special feature of the system at $\zeta_\infty=1$. At some fixed shell $r$, define the (unnormalized) probability density
\begin{equation}
    \rho_r(u) = N_r P(u|r)
\end{equation}
We use the experimentally-observed form $P(u|r)\sim e^{\alpha_r u}$ to get
\begin{equation}
    \rho_r(u)\sim N_r e^{\alpha_r u}\label{eq:density}
\end{equation}
Consider a small interval $[u,u+du]$. The bitstrings in that bin each have probability approximately $p\sim e^{-u}$. The number of bitstrings in the bin is $\rho_r(u)du$. Therefore, the total probability of observing one of the bitstrings in this interval is
\begin{equation}
    \begin{aligned}
        dM_r(u) &= \rho_r(u) e^{-u}du\\
        &\sim  N_r e^{(\alpha_r-1)u} du
    \end{aligned}
\end{equation}
We see that the exponent $\alpha_r=1$ is special because then every interval $du$ in $u$ carries roughly comparable total probability weight:
\begin{equation}
    \rho_r(u) e^{-u}\sim \text{const.}
\end{equation}
Integrating over $u$ to get the total probability mass,
\begin{equation}
    \begin{aligned}
        M_r &\sim  N_r \int_{u_\text{min}}^{u_\text{max}}du ~e^{(\alpha_r-1)u}\\
        &\sim N_r ~\frac{e^{(\alpha_r-1)u_\text{max}}-e^{(\alpha_r-1)u_\text{min}} }{\alpha_r-1}
    \end{aligned} 
\end{equation}
Suppose first that $\alpha_r>1$. Then the probability mass is dominated by $u_\text{max}$, giving
\begin{equation}
    M_r \sim N_r ~\frac{e^{|\alpha_r-1|u_\text{max}} }{|\alpha_r-1|} 
\end{equation}
This means that the probability mass is pushed towards the largest available $u$---in other words, towards the enormous population of indiviudally tiny-weight bitstrings (the ergodic limit). This cannot continue indefinitely, of course, because the total probability is normalized. For this reason we see the power-law scaling break down at small disorder, eventually giving way to Porter-Thomas statistics. Conversely, if $\alpha_r<1$, then the behaviour of $M_r$ is controlled by $u_\text{min}$:
\begin{equation}
    M_r \sim N_r ~\frac{e^{-|\alpha_r-1|u_\text{min}} }{|\alpha_r-1|} 
\end{equation}
Now the mass is biased towards smaller $u$---that is to say, bitstrings with large probability. In this regime, even though there are many low-probability bitstrings at larger $u$, their number does not grow fast enough to compensate their smaller probability. This corresponds to the `localized' or `frozen' limit. $\alpha_r=1$ is therefore the marginal exponent separating ``few dominant configurations'' from ``many configurations, each contributing with tiny probability''.
% \subsection{Multifractal dimension}
% The multifractal dimension $D_q$, quantifies how uneenly a normalized measure is distributed over a large set of configurations (in this case, basis states). It is defined as 
% \begin{equation}
%     \frac{1}{n^{D_q(q-1)}} = \sum_i p_i^{q}; \qquad p_i = |\psi_i|^2
% \end{equation}
% We proceed to evaluate this quantity assuming that $\mathcal{P}(p)$ follows a power-law distribution with exponent $1+\zeta$, as described in the main text:
% \begin{equation}
%     \begin{aligned}
%         \frac{1}{N^{D_q(q-1)}} &= \int dp  ~p^q \mathcal{P}(p)\\
%         &\propto \int_0^1 dp \frac{p^q}{p^{1+\zeta}}\\
%     \end{aligned}
% \end{equation}
% We first take care of normalization, using the fact that at $q=1$ we have $\sum_i p_i = 1$:
% \begin{equation}
%     \begin{aligned}
%         \int_0^1 \frac{dp}{p^{\zeta}} = \frac{1}{A}
%     \end{aligned}
% \end{equation}
% This gives $A = 1-\zeta$. Plugging this in,
% \begin{equation}
%     \begin{aligned}
%         \frac{1}{N^{D_q(q-1)}} &= (1-\zeta) \int_0^1 \frac{dp}{p^{1+\zeta-q}}
%     \end{aligned}
% \end{equation}
% Suppose first that $\zeta=1$. Then, it follows that 
% \begin{equation}
%     \int_0^1 \frac{dp}{p^{2-q}}
% \end{equation}

\vspace{200mm}

\newpage 
\section{Computing Binary Intrinsic Dimension}
\label{subsec:BID}

Consider a dataset consisting of binary variables with $N$ features. If these variables are independent, they should uniformly populate the vertices of a $N$-dimensional hypercube, with an intrinsic dimension of $N$. This information is contained in the distribution of Hamming distances $r$:
\begin{equation}
	P_0(r) = \frac{1}{2^{N}} \binom{N}{r}.
\end{equation}
The Binary Intrinsic Dimension (BID) algorithm~\cite{acevedo2025unsupervised} builds on two assumptions. The first one is that, if correlations occur, the data points can be effectively describes as if they lived on a lower-dimensional geometrical object. The second one states that the dimension $d$ of this object is a smooth function of the distance, $d(r)$. Hence, an ansatz is proposed for the distribution of hamming distances:
\begin{equation}
\label{eq:BID_ansatz}
	P(r) = \frac{C}{2^{d(r)}} \binom{d(r)}{r},
\end{equation}
where $C$ is a normalization constant. Empirically, one can see that the observed distribution of distances can be captured by performing a linear expansion $d(r)\simeq d_0 + d_1 r$, where $d_0$ represents the BID itself. Here, we obtain the parameters $d_0$ and $d_1$ by minimizing the Kullback-Leibler divergence between the empirical distribution of Hamming distances and the model in Eq.~\ref{eq:BID_ansatz}.

\newpage
\vspace{100mm}
\section{Qualitative theoretical description}
\label{sec:qual_theory}

\subsection{Short time dynamics of return probability}
\label{subsec:short_time}
This Subsection is intended to demonstrate that
the short time behavior of $R(t)$ at $Jt \leq 1$ is well-described by the simplest approximation of independent spin pairs, and to lay the foundation for the  more sophisticated approach developed in Sec.~\ref{subsec:spin_noise} and~\ref{subsec:exponent_eta}.
Within this approximation, the Hamiltonian is given by
\begin{equation}
H_0 = \sum_\mu H_\mu =  \sum_\mu \left[ J \tau^x_\mu + \varepsilon_\mu\tau_\mu^z \right]
\label{Hmu}
\end{equation}
where $\pmb{\tau}_\mu$ denote Pauli matrices of combined spin operators on the links: 
$\tau^x_\mu = S^+_rS^-_{r'} + h.c. $, $\tau^z_\mu = S^z_r - S^z_{r'}$,
while $ \mu \equiv \langle r, r' \rangle $ is defined on each lattice bond and
 $\varepsilon_\mu \equiv \varepsilon_{r,r'} = \frac12 (h_r - h_r')$.
 The meaning of approximation (\ref{Hmu}) is to account for  spin-flip events that occur at each
 link $\mu$ in the lowest order over coupling $J$; here we mean, in terms of original spins $\pmb{S}_r$, a
 projection of the two-spin Hilbert space (dimension 4) to the subspace $|\uparrow\downarrow\rangle,|\downarrow\uparrow\rangle$ only. From this Hamiltonian, one can calculate the return probability as a function of time. Because all spin flips are independent in this approximation, we solve for the dynamics of each pair, find the return probability for this pair and take a product over all pairs:
\begin{equation}
R(t) = \prod_{\langle r,r' \rangle}
\left[1 - \frac{J^2}{\varepsilon_{r,r'}^2 + J^2}\sin^2\left(t \,\sqrt{\varepsilon_{r,r'}^2 + J^2}  \right)\right] 
\label{short}
\end{equation}
where the product goes over all $2n$ pairs of nearest neighbors, here $\pmb{\tau}_\mu$ defined on each lattice bond $\langle r, r' \rangle \equiv \mu$. 
This approximation is good for $Jt \leq 1$, as one can observe from comparing the corresponding data in Fig.~\ref{fig:short_time}  with the calculation based on the Hamiltonian (\ref{Hmu}), see below. 
$R(t)$ defined by Eq.(\ref{short}) is a random quantity that depends on specific realization of 
$h_{\mathbf{r}}$. Representative $R(t)$ can be obtained by the averaging of $\ln R(t)$ over 
$h_{\mathbf{r}}$, since this is an additive quantity: $R_{typ}(t) = \exp{\overline{\ln R(t)}}$,
where $\overline{[...]}$ denotes averaging over independent random $h_r$.
Approximation of independent pairs (\ref{short}) leads to a finite  disorder-dependent 
limit for $R_{typ}(\infty)$ that can be calculated as
\begin{equation}
- \ln R_{typ}(\infty) 
\approx m\frac{4(\pi -2)}{w}
\label{Rinf}
\end{equation}
The $m \leq n$ factor counts the number of bonds on a square lattice with $n$ sites and with $h_r h_{r'} < 0$,
and the integral over the
distribution of $\varepsilon_{r,r'}$
is calculated to leading order in $1/w \ll 1$.

We numerically evaluate the expression for $R_{typ}(t)$ at $w=15$, plotting the results in Fig.~\ref{fig:short_time}a  by a blue dashed line (for $n=25$) and a grey dashed line, $n=36$. Results of direct numerical simulation of $R_{typ}(t)$ for $n=16,20$ and two instances of $n=25$ are shown by full lines, together with experimental data for $n=36$. Good agreement between approximation (\ref{short}) and the data is seen until $Jt \approx 1$ \, corresponding to $t \approx 30$ns, while at later time-scales qualitative deviations occur. Indeed, while the theoretical dashed curves shows damped oscillations approaching a constant value at $t \to \infty$, the data (in  brown ) demonstrate faster decay of the oscillation amplitude and overall downshift of $R(t)$, in agreement with Eq.(2) of the Main text. We will argue below that such a behavior is due to residual interaction between different spin pairs ${\pmb{\tau}}_\mu,\pmb{\tau}_\nu$ leading to dephasing and  inelastic transitions.

Power-law behavior  at long times means that the probability of flipping any spin $S_r$ in the array grows logarithmically with time like $P_\mathrm{{flip}} \sim (\eta/n) \ln (Jt)$; such behavior is akin to  spin noise with $1/f$ power spectrum, known to be generic for glasses and spin glasses~\cite{Esquinazi,Kogan,Noise}. Indeed, numerical simulations (see next Subsection) demonstrate a spin noise spectrum $\mathcal{S}(\omega) = \langle |S^z(\omega)|^2\rangle \approx \frac{A}{|\omega|}$.

\subsection{$1/f$ spin noise}
\label{subsec:spin_noise}
This Subsection demonstrates a relation between broad spectrum of relaxation rates (characteristic for usual glasses) and the  $1/f$ noise and its connection to the power-law decay of $R(t)$.
\begin{figure}[h]
    \centering
    \includegraphics[width=\linewidth]{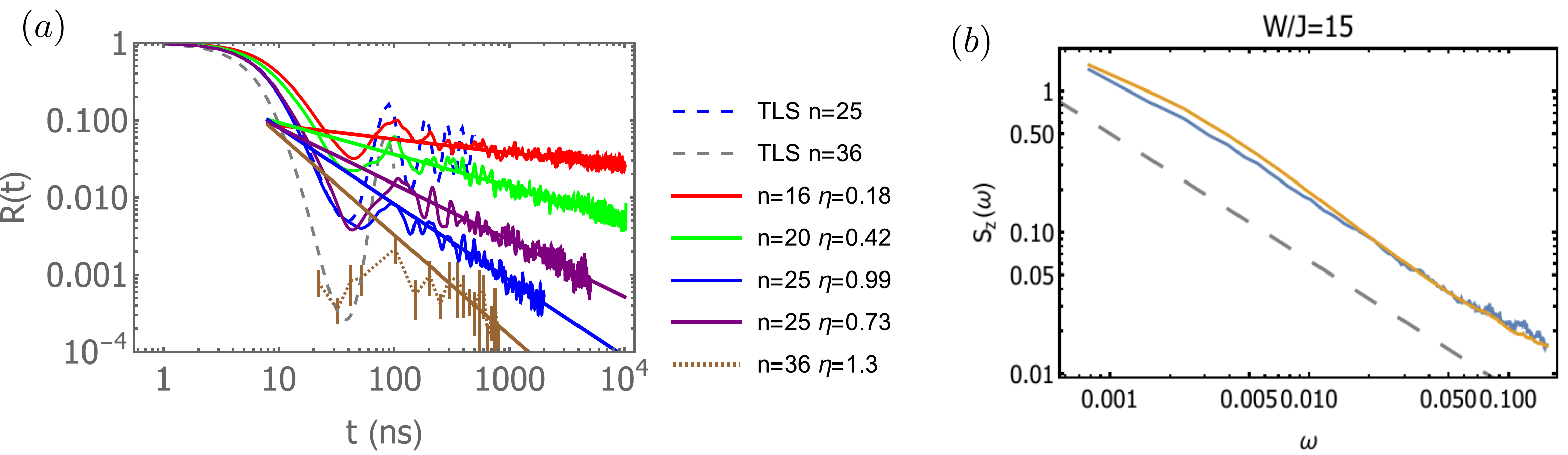}
    \caption{Panel a): Return probability $R(t)$ for moderate disorder $w=15$; dashed lines obtained with model Hamiltonian (\ref{Hmu}), full lines numerically for $n=16,20,25$, dotted line is for experimental data for $n=36$.  Panel b): Power spectrum of local spin $S^z(t)$ fluctuations: numerical simulation on the system of 
    $n=20$ spins with $w=15$.  }
    \label{fig:short_time}
\end{figure}
Slow relaxation in glasses or spin glasses is usually associated with the presence of  a broad spectrum $P(\Gamma) d\Gamma $
of relaxation rates  $\Gamma$ that characterize the dynamics of various modes of fluctuations, commonly known as two-level systems. If a broad distribution, $P(\Gamma) d\Gamma \sim d\Gamma/\Gamma$, is assumed, one immediately arrives 
~\cite{Noise} at the prediction of the noise spectrum  $\mathcal{S}(\omega) \propto 1/|\omega|$.
To check if such noise is present in our system, we performed numerical simulation of the dynamics described by \eqref{H1} for system size $n= 20$. We computed local spin-spin correlation functions 
$S_r(t) = \langle S_{\mathbf{r}}^z(0) S_{\mathbf{r}}^z(t) \rangle $ and then averaged over positions  $\mathbf{r}$ and realizations of disorder to obtain $S(t) = \overline{\langle S_r^z(0) S_r^z(t) \rangle}$.
Its Fourier transform $\mathcal{S}(\omega)$ is plotted in Fig.~\ref{fig:short_time}b. One observes $1/f$ noise over two decades in frequency.

\subsection{Exponent $\eta$ and its $n$-dependence.}
\label{subsec:exponent_eta}
Here we explain the power-law decay of $R(t)$ with exponent $\eta \propto n^2$, using
the phenomenology of two-level systems in glasses~\cite{TLS} and to the related mechanism of $1/f$ noise~\cite{Noise}, as discussed above in Sec.~\ref{subsec:spin_noise}.  The key point is the existence of a broad spectrum of very low frequencies in a system. The toy Hamiltonian described earlier (\ref{Hmu}) is not sufficient in that respect, since all its eigenvalues are $\geq J$.  Thus it is natural to take into account the interactions between effective spins $\pmb{\tau}_\mu$, which makes possible correlated flips of pairs of spins such that there is a small change in the total energy of a pair, see e.g.~\cite{BurinMirlin2016,FaoroIoffe2012}. Notice that the scaling described by \eqref{R-power}[Main text] cannot be valid at arbitrary large $n \to \infty$ since at any fixed large $Jt$ it would make $R(t)$  smaller than the minimal IPR value $I_2^{min} = 2^{-n}$.  However, due to the smallness of $\kappa (w) \leq 1/w^2$, 
%[see Eq.\eqref{eta} of the Main text], 
the upper bound $n_{max} \geq w^2$ compatible with such a dependence exceeds the maximal size  of our system, even for  critical disorder $w \approx 11$.  Thus our results for $\eta(n)$ dependence refer to the intermediate asymptotics of \textit{moderately large} systems with $1 \ll n \leq n_{max}(w)$.

Moving forward, it will be sufficient to consider only the spins satisfying $|\epsilon_\mu| \leq J$---we will call them ``coherent TLS". Assuming uniformly distributed disorder, the fraction of these spins is $\sim 2/w\ll1$. The interaction Hamiltonian between these pairs is given by
\begin{equation}
    H_{int} = \frac12\sum_{\mu\nu} \left(V_{\mu,\nu}^{ab}\,\tau_\mu^a\tau_\nu^b + h.c.\right)
    \label{TLS-int}
\end{equation}
so that the full Hamiltonian consists of Eqs.(\ref{Hmu},\ref{TLS-int}). The total number of effective spins participating in the dynamics, therefore, is approximately $n_* = 2n\times(2/w) \gg 1$.
Matrix elements $V_{\mu,\nu}^{ab}$ appear in higher orders of perturbation theory as powers of $j=1/w \ll 1$: these represent interactions between rare active TLS's via the ``inert media" of all other spins. Therefore,  $V_{\mu,\nu}^{ab}$ are mostly  small compared to $J $,
and the statistical distribution of their absolute values is very broad. Writing $|V_{\mu,\nu}^{ab}| \sim J \exp(- L_{\mu,\nu})$ we present this distribution via
\begin{equation}
\mathcal{P}(L)\,dL \approx P_1 \, dL
\quad \mathrm{where} \quad L_{\mathrm{min}} < L < L_{\mathrm{max}}
\label{PL}
\end{equation}
with  
$L_{\mathrm{min}} \sim \ln w$ and  $ L_{\mathrm{max}} \sim n_* \ln w \gg L_{\mathrm{min}}$.

Now we formulate the key assumption of the further theoretical analysis: the system of interacting TLS's is characterized (for values of $w$ corresponding to the glassy state) 
by a dephasing rate $\Gamma_\phi$ that is much larger than the relaxation rate $\Gamma$.
Theoretical arguments demonstrating the possibility of such a situation are provided in Sec.~\ref{sec:daphasing_relaxation} below. 
Qualitatively, spin dephasing
without spin relaxation is equivalent to the presence of energy transport in the absence of full ergodicity; low-temperature glasses provide a useful example of such a situation. Indeed, dephasing-only is present as soon as the effective magnetic field $h_r^z$ acting at a site $r$ 
acquires a slow time dependence due to the coupling of $\pmb{S}_r$ to other spins. Since the total energy of the system is conserved, this is possible if energy transport is allowed between different parts of the system. Within the MBL framework, studies of energy transport were provided in Ref.~\cite{EnergyDiffusion-Varma,EnergyDiffusion-Prelovcek}.

We expect $\Gamma_\phi \ll J$ but still larger than most of the matrix elements $V_{\mu,\nu}$ since the latter are due to high-order processes. Then \textit{real transitions} leading both to $1/f$ noise and to logarithmic growth of $-\ln R(t)$  occur due to the interaction (\ref{TLS-int}) between pairs of active TLS
with small $|E_\mu - E_\nu|$, where $E_\mu = \sqrt{\varepsilon_\mu^2 + J^2}$ is the energy splitting for the non-interacting model. An estimate for the rate $r_{\mu\nu}$ of such (incoherent) transition is given by $r_{\mu\nu} \sim V_{\mu,\nu}^2 \Gamma_\phi/|E_\mu - E_\nu|^2$\, (compare with Eq.(26) from
Ref.~\cite{Noise}). Exponentially broad distribution of matrix elements $V_{\mu,\nu}$ defined by Eq.(\ref{PL}) translates then to
the same kind of distribution for the rates $ r_{\mu\nu} \sim \Gamma_\phi e^{-2L_{\mu\nu}}$. Factor $P_1$ in Eq.(\ref{PL}) is proportional to $n$ due to the normalization condition $\int \mathcal{P}(L) dL = P_1 L_\mathrm{{max}} \propto n_*^2$ since total number of pairs of interacting TLS scales as $n_*^2$, and also $L_{\mathrm{max}} \propto n_*$. In result, we find that $P_1 = \gamma \, n$ where $\gamma \equiv \gamma(w)$. 

Probability for any spin to be involved in inelastic relaxation during long time $t \gg 1/J$ grows 
$\propto \gamma\, n \ln(Jt)$.  Now, in order to obtain the estimate for the full probability of return $R(t)$, one needs to take into account that $\ln R(t)$ is additive over all spins, thus an additional multiplication on the number of spins $n$ should be performed. This way one gets  power-law decay, $ R_{\mathrm{typ}}(t) \sim (Jt)^{-\eta}$, 
with exponent $\eta \propto n^2$, not far from its experimentally observed dependence
in Eq.(4) of the Main text.

\subsection{Dephasing without relaxation: results of a model calculation}
\label{subsec:dephasing_without}

The analysis in Sec.~\ref{subsec:exponent_eta} above
explains  the power-law dependence $R(t) \sim 1/t^\eta$ with exponent $\eta(n) \propto n^2$, 
using \textit{the assumption} of  dephasing rate $\Gamma^\phi $ being much higher than rate of inelastic relaxation 
$\Gamma^r$.  Below in Sec.~\ref{sec:daphasing_relaxation} we present a  calculation for the Hamiltonian (1) of the Main text, which
demonstrate the existence  of intermediate range of disorder, $w_r < w < w_\phi$, where lowest-order spin 
relaxation processes are absent, while spin dephasing is active. Here in this Subsection we provide a brief account of that calculation, concentrating on its main results.

%In fact, we assumed that non-zero $\Gamma^\phi >0 $ is generated at $ j_c < J/W \ll 1$ once original couplings $J$ between  neighboring spins in the Hamiltonian (1) [Main text] are taken into account, while for relaxation (real spin flips) an analogous 
%critical value $j_r = J_r/W$ is  larger, $j_r > j_c$.  In such a case at $j_c < J/W < j_r$ relaxation processes are due to high-order terms of the $J/W$ expansion, thus the inequality 
%$\Gamma^r \ll \Gamma^\phi$ holds. Below in Sec.4 we present a  calculation for the Hamiltonian (1) of the Main text,  but living  on the Cayley tree with branching number $K \gg 1$, which demonstrates  $j_c \ll j_r$.  Here in this Subsection we provide a brief account of that calculation, concentrating on its main results.

We define $\Gamma^\phi$ as $\Im\Sigma^\pm$ where $\Sigma^\pm(\omega)$ is the self-energy part corresponding to transverse spin Green
function $G^\pm(\omega)$; similarly, relaxation rate $\Gamma^r$ is defined via self-energy of longitudinal Green function $G^{zz}(\omega)$. Using Heisenberg equations for spin operators 
%and using the absence of loops on Cayley tree, 
one can derive, within self-consistent approximation (see next Sec.~\ref{sec:daphasing_relaxation})  recursion relations for local relaxation rates $\Gamma^r_i$, similar (but different) to those of Ref.\cite{FIM2010}:
\begin{equation}
\Gamma_0^r = J^2 \sum_{j} \frac{\Gamma_j^r}{(h_j-h_0)^2 }
\label{Gammar}
\end{equation}
Summation here goes over $K$ neighbors of the "central spin"  $\pmb{S}_0$.
Below we use bare distribution of local energies $P_0(h) = \frac{1}{W}\theta (W/2 - |h|)$.
Recursions (\ref{Gammar}) converge under iterations
to $\Gamma^r =0$ at $w > w_r$, while at smaller $w$ linear iterations diverge and nonlinear in $\Gamma^r$ 
terms are needed to get  a final nonzero result.  Next, we assume $w > w_r$ and
derive (in the next order over $J^2$) analogous equations for $\Gamma^\phi$ :
\begin{equation}
\Gamma_0^\phi = J^4 \sum_{j\neq k}\left( \frac1{h_0-h_j} + \frac1{h_0-h_k} \right)^2 \frac{\Gamma_j^\phi + 
\Gamma_k^\phi}{(h_j-h_k)^2}
\label{Gamma0}
\end{equation}
Summation goes over  $K(K-1)/2$ pairs of unequal neighbors of the site $0$.  
Eqs.(\ref{Gamma0}) can be used to determine the boundary of the  parameter region where non-zero 
$\Gamma^\phi$ are generated, that is $J_\phi = W/w_\phi$. In general, equations for $\Gamma^\phi$ contain also terms of order $J^2$, but these terms are inefficient at $J < J_r$ and thus omitted here.

At very large $w$ we can neglect correlations between $\Gamma_i^r$ at some site $i$ and local field $h_i$; it amounts to setting
$h_0 =0$ in Eq.(\ref{Gammar}). Then these equations become identical to those derived in Ref.~\cite{AAT} for Anderson localization on a tree (of  branching number $K$), within simplest "upper limit" approximation that neglects level repulsion  (that is, real part of self-energy).
Then critical value  $w_r$ obeys simple equation
$w_r/2 = eK \ln(w_r/2)$. Its solution for $K=3$ is $w_r \approx 53.7$.
Generalizing the same procedure for recursion equations (\ref{Gamma0}), one can find (see Sec.~\ref{sec:daphasing_relaxation} below) the
critical value $w_\phi$ within "upper limit approximation". It is equal to $w_\phi \approx 71$, thus confirming that $w_\phi > w_r$. The same inequality is valid for larger values of $K$.

One can make a better calculation for both $w_r$ and $w_\phi$ taking into account corrections to the real part of self-energy function. For Eq.(\ref{Gammar}) we can  follow closely the approach proposed in Ref.~\cite{AKI2018} and get $w_r \approx 33$ for $K=3$. Generalization of the same approach to the case of recursions~(\ref{Gamma0}) is provided in Sec.~\ref{sec:daphasing_relaxation} below; it leads to $w_c \approx 50$ at $K=3$. The ratio $w_\phi/w_r \approx 1.5$ was found for larger values of $K$ as well.
We conclude that various approximation schemes consistently support our qualitative picture: in some range of $w$ values  the single-spin dephasing rate is much stronger than its relaxation rate.

\subsection{Exponent $\zeta$ in the strong-disorder limit}
\label{subsec:exponent_zeta}

Data  in Fig.~\ref{WavefuncStats}c and \ref{WavefuncStats}d [Main text] show: 
at strong disorder the distribution function 
$\mathcal{P}(\ln(\bar{\Psi}\Psi)) \sim (\bar{\Psi}\Psi)^{-\zeta} $, with the exponent $\zeta \to 1$ in the
limit of very large $w > 35$. We demonstrate now that such a behavior can be understood as the result of independent dynamics of "composite spins"   $\pmb{\tau}_\mu$, see Eq.(\ref{Hmu}) and definitions below it. 
Flip of any $\pmb{\tau}_\mu$ variable describes motion of hard-core boson from one site of the link $\mu$
to another one. Below it will be more convenient to work with the probability density $\mathcal{P}_1(p)$ for occupation probabilities $p=\bar\Psi\Psi$ themselves, thus the above definition of the exponent $\zeta$
corresponds now to $\mathcal{P}_1(p) \sim 1/p^{1+\zeta}$. 

\begin{figure}[h]
    \centering
    \includegraphics{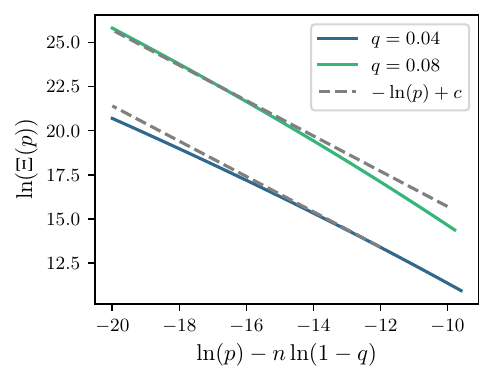}
    \caption{ The function $\ln(\Xi(p))$ versus $\ln(p)$ for $n = 100$ and various small $q$. Grey dashed lines
    are linear in $\ln(p)$ with slope (-1). We consider the range where $y_p > 3$ only,  to demonstrate the regime where Eq.~(\ref{P113}) is applicable.}
    \label{fig:Xi}
\end{figure}

Consider first just a single link $\mu \equiv (r,r')$ prepared in the state $|0\rangle = (1,0)$. The probabilities to find it
in states $|0\rangle$ and $|1\rangle = (0,1)$ at time $t$ are given by
\begin{equation}
p_0(t) = 1 - \frac{J^2}{\varepsilon_\mu^2 + J^2}\sin^2\left(\sqrt{\varepsilon_\mu^2 + J^2}\, t\right), 
    \quad 
    p_1(t) = \frac{J^2}{\varepsilon_\mu^2 + J^2}\sin^2\left(\sqrt{\varepsilon_\mu^2 + J^2} \, t\right)
    \label{p0}
\end{equation}
%where we have introduced $\Omega \equiv \sqrt{\varepsilon_\mu^2 + J^2}$. 
The probability distribution
over probabilities $p$ is then given by
\begin{equation}
    \mathcal{P}_1^{(1)}(p) = \frac{1}{2}\delta(p - p_0(t)) + \frac{1}{2}\delta(p - p_1(t))
    \label{P10}
\end{equation}
Since the values of $\varepsilon_\mu$ are broadly distributed with width $2W$, in nearly
all cases the magnitudes of $p_1(t)$ are very small.  However, the total number of different 
links $\mu$ is large, as it is proportional to $n$. Let us consider a simple model where for each link the probability of finding the system in the state $|1\rangle$ is $p_1(t) = q \ll 1$, and calculate
the probability density $\mathcal{P}_1^{(n)}(p)$ under the assumption that $|\ln(p)| \ll n$:
\begin{equation}
    \mathcal{P}_1(p) = \frac{1}{2^n}\sum_{k = 0}^n \binom{n}{k} \, \delta\!\left(p - q^k (1 - q)^{n-k}\right).
    \label{P112}
\end{equation}
Assuming the dominant contribution to the above sum comes from such values of $k$ than  $1 \ll k \ll n$,
we rewrite the sum via the following integral:
\begin{equation}
    \mathcal{P}_1(p) = \frac{1}{2^n}\int_0^{\infty} dy \, \frac{n^y}{\Gamma(y+1)} \,
    \delta\!\left(p - q^y (1 - q)^{n-y}\right) 
    = \frac{n^{y_p}}{2^n \Gamma(y_p + 1)} \,
    \frac{1}{\log(q) - \log(1 - q)} \, \frac{1}{p} \equiv \frac{\Xi(p)}{2^n p}
    \label{P113}
\end{equation}
where $y_p$ is defined implicitly by the relation
\begin{equation}
    p = q^{y_p} (1 - q)^{n - y_p}
    \label{yp1}
\end{equation}
The function $\Xi(p)$ is defined above in Eqs.(\ref{P113},\ref{yp1}), and for various small values of 
$q \ll 1$ it can be found numerically. The result of this computation is shown in Fig.~\ref{fig:Xi}
for $q=0.04$ and $q=0.08$. Clearly, $\Xi(p) \sim 1/p$ for a broad range of moderately small values of $p$,
confirming therefore that $\mathcal{P}_1 \propto 1/p^2$ within the model of dynamics of independent links,
Eq.(\ref{p0},\ref{P10},\ref{P112}).

Interaction term (\ref{TLS-int}) is responsible for a  collective dynamics of  $\pmb{\tau}_\mu$   variables, 
leading to faster decay of $\mathcal{P}_1 \propto 1/p^{1+\zeta}$, corresponding to $\zeta > 1$.
Experimental results shown in  Fig.~\ref{WavefuncStats}[Main text] are consistent with asymptotic approach of $\zeta$ to unity
as $w \to \infty$, in the same way as exponent $\eta$ approaches zero in the same limit, see Fig.~\ref{R(t)}.e[Main text].

\newpage
\section{Relaxation and dephasing  in spin-$\frac12$ model on the Cayley tree}

\label{sec:daphasing_relaxation}

The main goal of the theory developed below is to demonstrate that the local nature of spin-spin interaction in a random system indeed lead to the  spin dephasing time  much shorter than spin relaxation time - to support
the hypothesis we used above, in Sec.~\ref{subsec:exponent_eta} and~\ref{subsec:dephasing_without}. We will use a simplification of our problem; namely, we consider spins $\frac12$ siting on a Cayley tree (infinite Bethe lattice) instead of a real square lattice. 
Cayley tree models serve usually as a mean-field approximation for the real problems formulated in Euclidean space,
the key simplification being the absence of closed loops on a tree. Moreover, for the problems with strong randomness (like we are dealing with), loop-less approximation looks especially reasonable. Detailed argumentation for that can be found, for example, in Ref.~\cite{Yan1992,Khvalyuk2024}.
The key idea behind  is however rather simple: strong local disorder makes couplings between nearby sites of 
the lattice very much unequal, suppressing efficient "communications" between the sites along closed loops.
Eventually, the Euclidean structure of the space will prevail at very large sizes, but the presence of a relatively large disorder-induced intermediate scale enhances reliability of tree-like approximations. The same reasoning explains our observation [Main text] of an intermediate scaling $\eta \propto n^{2.4}$ within a broad range of system sizes until $n=70$ (and possible much above), although eventually \textit{in thermodynamic limit} $n \to \infty$ the crossover to linear scaling of $\eta(n)$ is expected due to locality in real space.

\subsection{Preliminaries}
\label{subsec:preliminaries_relaxation}
We are going to study a system of spins-$\frac12$ in a random magnetic field with XY interaction on a Cayley tree (CT)
 with large branching number  $K \gg 1$. The Hamiltonian is of the following form:
\begin{equation}
    H = -J \sum_{\langle i, j\rangle} (\sigma^+_{i} \sigma^-_{j} + 
\sigma^+_{ j^\prime}  \sigma^-_i ) + \frac12 \sum_{i} h_i \sigma_{i}^z.
\label{S1}
\end{equation}
where $i,j$ are sites the Cayley tree, summation in the first term goes over links $\langle i, j\rangle$ of nearest 
neighbours. Random variables $h_i$ belong to uncorrelated symmetric box distribution of full width $W$.
 Operators $\sigma^a_i = 2S^a_i$ with $a=1,2,3$ are Pauli matrices, while $\sigma^\pm_i \equiv S^\pm_i$.
Thus Hamiltonian (\ref{S1}) is equivalent to the Hamiltonian (1) from the Main text up to the replacement of square
lattice by the Cayley tree. Full coordination number of sites of CT is $Z=K+1$ and we have in mind correspondence to
original square lattice, so the "most realistic" value of branching number is $K=3$.

Our main goal is to derive, using loopless structure of the Cayley tree, a kind of recursion relations between i) local
relaxation rates $\Gamma_i^{r}$ for spins which belong to different sites $i$ of the CT, and ii) local dephasing rates
$\Gamma_i^\phi$ for the same spins. To realize this program, it will be useful to represent Heisenberg equations for spin
operators in the form similar to Schrodinger equations for wavefunctions. The advantage of this approach is that it
keeps track of local nature of interactions in the Hamiltonian (\ref{S1}), which is somewhat hidden in  many-body wavefunctions.

The evolution equation for the operators is a linear equation and to rewrite it as a Schr\"odinger-type equation 
 we introduce the mixed state of our system with its copy:
\begin{equation}
    |\hat{1}\rangle := \bigotimes_{j} \left[\frac{|00\rangle_j + |11 \rangle_j} {\sqrt{2}}\right] = \frac{1}{\sqrt{2^n}}\sum_{s} |s\rangle \otimes |s\rangle
\end{equation}
where $n$ is the total number of spins.
In the last terms sum runs over all bit-strings.  Using this mixed state we can consider  each operator as a wave function in the extended Hilbert space (of original system and its copy):
\begin{equation}
    |\hat{O}\rangle = \hat{O} \otimes \hat{1} |\hat{1}\rangle
    \label{OperatorWf}
\end{equation}
We note the following useful property: $\hat{O} \otimes \hat{1} |\hat{1}\rangle =  \hat{1} \otimes \hat{O}^T |\hat{1}\rangle$.
The evolution equation acquires then the following form: 
\begin{eqnarray}
 \label{S2}
    i\partial_t |\hat{O}(t) \rangle & = & \mathcal{H} |\hat{O}(t) \rangle \equiv  |[\hat{O}(t),H] \rangle  \quad \mathrm{where} \\ \nonumber
     \mathcal{H} & = &  \hat{1}  \otimes H^T - H \otimes \hat{1} 
\end{eqnarray}
Below we consider Hamiltonian (\ref{S1}) as a sum $H = H_0 + V$ where "bare Hamiltonian" $H_0$ coincides with the second sum
in (\ref{S1}) while "perturbation" $V$ is given by the first term.  
The states $|\sigma_j^z\rangle$,  $\sqrt{2}|\sigma_j^\pm\rangle$ (in the extended Hilbert space) 
are  normalized eigenstates of $\mathcal{H}_0$ with eigenvalues $0$ and $\pm h_j$ respectively.

The action of the perturbation $\mathcal{V}$ on the operators $\sigma^z_j$ and $\sigma_j^\pm$ is as follows:
\begin{eqnarray}
    \mathcal{V}|\sigma_j^z\rangle & = & J \sum_{k \in \partial j} |[\sigma^z,\sigma_j^+ \sigma_k^- + \sigma_j^- \sigma^+_k ]\rangle  \\ \nonumber
    & = & 2 J \sum_{k \in \partial j}\left( |\sigma_j^+ \sigma_k^-\rangle  - |\sigma_j^- \sigma^+_k \rangle\right) 
    \label{Vz}
\end{eqnarray}
and
\begin{equation}
     \mathcal{V}|\sigma_j^\pm \rangle  = \pm J  \sum_{k \in \partial j}|
    \sigma^z_j  \sigma^\pm_k\rangle
    \label{Vpm}
\end{equation}
Here $k\in \partial j$ means that $k$ is a neighbor of $j$.

Now we  calculate relaxation rate $\Gamma^\pm$ in this spin model. 
The relaxation rate is given by the imaginary part of the self-energy $\Sigma_z$ which reads:
\begin{equation}
    \Sigma_z(\epsilon) = \Im \langle \sigma_j^z|\mathcal{V}\left(\epsilon - \mathcal{H} + i0 \right)^{-1}\mathcal{V}|\sigma_j^z\rangle
\end{equation}
 
Now we use approximation of  high connectivity, $K \gg 1$. It allows us to consider dynamics of neighboring spins 
being weakly correlated, thus $\langle \sigma^+_j(t) \sigma^+_k(t)|\sigma^+_j \sigma^+_k\rangle \approx  \langle \sigma^+_j(t) |\sigma^+_j\rangle \langle \sigma^+_k(t)|\sigma^+_k\rangle$.  This  is a kind of self-consistent Born approximation, like the one
used in Ref.~\cite{BAA2006}. 
The resulting self-energy for $z$-component of spin is given by:
\begin{eqnarray}
\label{Sigmaz}
    \Sigma_z^{(j)}(\epsilon) = 2 J^2 \sum_{k \in \partial } \left[\langle \sigma_k^-|(\epsilon - h_j - \mathcal{H} + i0)^{-1}|\sigma_k^-\rangle + \right. \nonumber\\ \left. \langle \sigma_k^+|(\epsilon + h_j - \mathcal{H} + i0)^{-1}|\sigma_k^+\rangle\right] \nonumber\\
\end{eqnarray}
In the same way, self-energy for $\sigma^\pm$ components can be found in the form
\begin{equation}
\Sigma_\pm^{(j)}(\epsilon) =  2 J^2 \sum_{k \in \partial j} \langle \sigma_k^\pm|(\epsilon - \mathcal{H} + i0)^{-1}|\sigma^\pm_k\rangle
\label{Sigmapm}
\end{equation}

\subsection{Recursion equations for the relaxation rates}
\label{subsec:recursion_equation}
Relaxation rate $\Gamma_j$ can be found either via the imaginary part of $\Sigma_z^{(j)}(\epsilon)$ at very low energy $\epsilon  \to 0$
or (if there are no additional channels of dephasing) via the imaginary part of  $\Sigma_\pm^{(j)}(\epsilon)$ at $\epsilon= h_j$. Calculations of $\Im\Sigma$ using Eq.(\ref{Sigmaz}) in the first case, or Eq.(\ref{Sigmapm}) in the second case, lead to identical results (assuming that pure dephasing is absent), 
so we obtain the following  self-consisting equation for $\Gamma^{\pm}_j$ (factor 2 from RHS of  Eq.(\ref{Sigmapm}) cancels due to
normalization condition for $\sqrt{2}|\sigma_j^\pm\rangle$):
\begin{equation}
    \Gamma_j^{\pm} = J^2 \sum_{k \in \partial j} \frac{\Gamma_k^{\pm}}{(h_j - h_k)^2 + (\Gamma_k^{\pm})^2 }
    \label{SGammar}
\end{equation}
Summation in Eq.(\ref{SGammar}) goes over $K$ descendants of the site $j$. Below  we will find critical magnitude of disorder $w_r$ such that at $J > W/w_r$ 
recursions (\ref{SGammar}) lead to nonzero relaxation rates $\Gamma_j^{\pm}$.

\subsection{Calculation of dephasing rates}
\label{subsec:calcualtion_of_dephasing}
\subsubsection{Schrieffer - Wolff transformation}

Now suppose we consider range of couplings $J < J_r$ where lowest-order perturbation theory over  $J/W$ does not lead to relaxation
and dephasing; in other terms, recursion relations (\ref{SGammar}) leads to trivial solution with vanishing $\Gamma^\pm$. 
We are going to show that in fact this result does not mean that dephasing is really absent in our system.
To demonstrate it, we study second-order processes, which may lead to \textit{pure dephasing} without relaxation.

It is convenient to employ Schrieffer - Wolff unitary transformation generated by anti-hermitian operator $U$.
 
Then the effective Hamiltonian becomes of the form
\begin{equation}
    H_{eff} = H + [H,U] + \frac{1}{2}[[H,U],U] + \ldots
\end{equation}
We are going to find operator $U$ such that $H_{eff}$ will not contain terms which are linear in $J$ and proportional 
 $\sigma_j^\pm$. 
Let $U = \sum_{\langle j,k \rangle}s_{jk}(\sigma_j^+ \sigma_k^- - \sigma_j^- \sigma_k^+)$. Then we need to obey
\begin{equation}
    [H_0,U] + V = 0.
\end{equation}
The above equation gives: $s_{jk} = J/(h_j - h_k)$. Then the effective Hamiltonian is $H_{eff} = H_0 + V_{eff} $ where
$V_{eff} = \frac{1}{2}[V,U] $. Calculating the commutator, we find
\begin{align}
     V_{eff} =
   \frac{J}{2}\sum_{\langle j,k\rangle}\sum_{\langle l, m\rangle}s_{jk}[\sigma^+_l \sigma^-_m + \sigma^+_l \sigma^-_m,\sigma^+_j \sigma^-_k - \sigma^+_k \sigma^-_j]
   \label{Veff}
\end{align}
The commutator in Eq.(\ref{Veff})  is not vanishing if the edges $(j,k)$ and $(l,m)$ coincide  or have at least 
one common vertex. 
Thus we can rewrite $V_{eff}$ as follows:
\begin{align}
    V_{eff} =   \frac{J}{2} \sum_{\langle j,k\rangle}s_{jk}(\sigma^z_k - \sigma^z_j) - 
    \frac{J}{2} \sum_j \sum_{(l,k)\in \partial j}(s_{jk} + s_{jl})\sigma^z_j (\sigma^+_l \sigma^-_k +  \sigma^-_l  \sigma^+_k)
    \label{Veff2}
\end{align}
Here sign $(l,k)\in \partial j$ means that the sum runs over all possible pairs of neighbors of $j$-th vertex. 
Effective perturbation Hamiltonian (\ref{Veff2}) will be used below to calculate dephasing rates.

\subsubsection{Recursion equations for dephasing}

We will use the formalism of the previous Subsection to find a dephasing rate, with a focus on dynamics of operator $\sigma^+_j$.  
The action of the perturbation $V_{eff}$ on a $\sigma^+_j$ can be written, in the extended space, in the form
\begin{equation}
    \mathcal{V}_{eff}|\sigma^+_j\rangle \approx  J \sum_{(m,k)\in \partial j}(s_{jk} + s_{jm})|\sigma^+_j(\sigma_k^+ \sigma_m^- + \sigma_k^- \sigma_m^+ ) \rangle
\end{equation}
Note, that we take into account the terms  proportional to $\sigma^z_j$ only, 
while other terms which are usually  responsible for  decay process are neglected; the reason is our assumption that
$J/W$ is too small to generate decay self-consistently. 

The corresponding self-energy is:
\begin{align}
\Sigma_j^+(\epsilon) =4
J^2 \sum_{(m,k)\in \partial j}(s_{jk} + s_{jm})^2 \langle\sigma_k^+ \sigma_m^-| (\epsilon - h_j - \mathcal{H}_{eff} + i0)^{-1}|\sigma_k^+ \sigma_m^-\rangle 
%\approx \nonumber\\
%4
%J^2 \sum_{(m,k)\in \partial j}(s_{jk} + s_{jm})^2  \langle\sigma_k^+ \sigma_m^-| (\epsilon - h_j - \mathcal{H}_{eff} + i0)^{-1}|%\sigma_k^+ \sigma_m^-\rangle
\end{align}
The dephasing rate $\Gamma^\phi$ is given by the imaginary  part of the self-energy $\Im\Sigma_j^+(\epsilon)$ 
at $\epsilon  = h_j$; note that this case is opposite to the one considered in Sec.~\ref{subsec:recursion_equation} where 
all dephasing was due to spin relaxation. 
To estimate a correlation function of the operator $\sigma^+_k \sigma_m^-$ we use large connectivity limit $K \gg 1$ and suppose that this  correlation function is factorized  in the  time domain, due to weak correlations between different spins; the same approximation
was used in the previous Section while deriving Eq.(\ref{Sigmaz},\ref{Sigmapm}). 
The resulting equation for the self-energy reads as follows:
\begin{align}
   \Sigma_j^+(h_j) = 4J^2 \sum_{(m,k)\in \partial j}(s_{jk} + s_{jm})^2 i\int \frac{d\epsilon_1}{2\pi} \langle\sigma_k^+| (\epsilon_1 - \mathcal{H}_{eff} + i0)^{-1}|\sigma_k^+ \rangle  \times \nonumber\\ \langle \sigma_m^-| (- \epsilon_1 - \mathcal{H}_{eff} + i0)^{-1}| \sigma_m^-\rangle \approx 
   J^2 \sum_{(m,k)\in \partial j} \frac{(s_{jk} + s_{jm})^2}{h_m - h_k + i (\Gamma^{\phi}_k + \Gamma^{\phi}_m) }
\end{align}
Calculating the imaginary part of the above self-energy we find the (linearized form of) recursion equation for the dephasing rate:
\begin{equation}
    \Gamma_{j}^{\phi} = J^4\sum_{(m,k)\in \partial j}\left(\frac{1}{h_j - h_k  } + \frac{1}{h_j - h_m  } \right)^2 \frac{\Gamma^{\phi}_k + \Gamma^{\phi}_m}{(h_k - h_m)^2}
    \label{SGammaphi}
\end{equation}
Recursion equation (\ref{SGammaphi}) will be used below  for determination of the threshold value
$J_c = W/w_c$ for the existence of self-consistent dephasing in our spin system.

%\newpage
\subsection{Thresholds for the relaxation and dephasing channels}
\label{subsec:thresholds}

\subsubsection{``Upper limit" approximation}

\vspace{0.5cm}
\hspace{1cm}\textit{1). Relaxation rate}
\vspace{0.5cm}

At very small $J/W$ we can neglect correlations between $\Gamma_i^r$ at some site $i$ and local field $h_i$; it amounts to setting
$h_0 =0$ in Eq.(\ref{SGammar}). Then these equations become identical to those derived in Ref.~\cite{AAT} for Anderson localization on a tree (of  branching number $K$), within simplest "upper limit" approximation that neglects level repulsion 
(in other terms, real part of self-energy is neglected).
The critical value  $w_r = W/J_r$ can be found from the analysis of the linearized version of Eq.(\ref{SGammar}). Instability point
of this linear recursion (with $\Gamma^\pm$ neglected in denominator in R.H.S.) is determined by the set of two equations
\begin{eqnarray}
\label{crit}
\frac{d F}{d x}\left|_{x=x_*}\right. = 0 \, ; \qquad
F(x_*) = 0
\end{eqnarray}
where $F(x) \to F_1(x)$ and
\begin{eqnarray}
\label{F1}
F_1(x) & = & \frac{1}{x}\ln\left\{K \int P_0(h) dh \left[J^2\cdot f(h) \right]^x\right\} ; ~~~~~~~ f(h) = 1/h^2
\end{eqnarray}
where $P_0(h) = (1/W)\theta(\frac{W}{2} - |h|)$.
Equations (\ref{crit},\ref{F1}) can be derived using theory of freezing transition for random polymers on a tree~\cite{polymer1,polymer2}.
Solving the above set of equations one finds algebraic equation
\begin{equation}
\frac{w_r}{2} = e K \ln\frac{w_r}{2}
\label{Jr}
\end{equation}
Its solution for $K=3$ is $w_r \approx 53.7$.

\vspace{0.5cm}
\hspace{1cm}\textit{2). Dephasing rate}
\vspace{0.5cm}

Generalizing the procedure described in the above Subsection for recursion equations (\ref{SGammaphi}), 
one can find critical value $w_\phi = W/J_\phi$ by means
of the same Eqs.(\ref{crit}) but with $F(x)$ function replaced by another function $F_2(x)$ where
\begin{equation}
F_2(x) = \frac{1}{x}\ln\left\{ K^2\iint_{-1/2}^{1/2} dh_1 dh_2   \left[ J^4\cdot f(h_1,h_2)\right]^x\right\}
\label{F2}
\end{equation}
where 
\begin{equation}
f(h_1,h_2) = \frac{1}{(h_1-h_2)^2}\left(\frac{1}{h_1} + \frac{1}{h_2} \right)^2
\label{fx}
\end{equation}

Equations (\ref{F2},\ref{fx}) can be derived along the  line of ideas present in Refs.~\cite{polymer1,polymer2}.
We introduce Laplace transform $G(e^{-x})$ of the probability density function $\mathcal{P}(\Gamma_j)$ where evolution
of $\Gamma_j$ along the recursion follows Eq.(\ref{SGammaphi}). Then, instead of Eq.(9) of Ref.~\cite{polymer2} we come to
\begin{equation}
    G_{L+1}(x) = \prod_{\mu=1}^{K(K-1)/2} \int \rho(\epsilon_\mu) d\epsilon_\mu G_L^2(x + \epsilon_\mu) 
    \label{recursion2}
\end{equation}
where $\mu$ stays for a pair of indices $j,m$ staying in the R.H.S. of Eq.(\ref{SGammaphi}), number of such pairs is
$K(K-1)/2$. Density of the distribution $\rho(\epsilon)d\epsilon$ is determined by the original distribution of local 
fields $P_0(h_1,h_2) = W^{-2}\theta(\frac{W}{2}-|h_1|)\theta(\frac{W}{2}-|h_2|)$ and by the function (\ref{fx})
that corresponds to the $h$-dependent factor in Eq.(\ref{SGammaphi}). We set $h_j \to 0$ to estimate the critical values of the disorder.  %after we set $h_j \to 0$, as it was done in the beginning of Sec. II A 1 above. 
Second power of the characteristic function $G$ comes about in Eq.(\ref{recursion2})  due to the presence of two random variables $\Gamma_k,\Gamma_m$ in the R.H.S. of Eq.(\ref{SGammaphi}).
General properties of the characteristic function $G(x)$ are the same as in the original approach~\cite{polymer1,polymer2}.
In particular, freezing transition point is determined by its far right asymptotics $x \to \infty$ where 
$1 - G(x) \equiv g(x) \ll 1$, and thus $G^2(x) \approx 1 - 2g(x)$. Due to that fact, we come to Eq.(\ref{F2}),
where we also replaced $K(K-1)\to K^2$ in the large-$K$ limit.

Approximate integration in Eq.(\ref{F2}) and use of (\ref{crit}) gives
\begin{equation}
\frac{w_\phi}{(2\sqrt{2} + 2)^{1/2}} = e K \ln (w_\phi\sqrt{\alpha_1})
\label{Jphi}
\end{equation}
where $\alpha_1 \sim 1$. It follows from Eqs.(\ref{Jphi},\ref{Jr}) that $w_\phi > w_r$ in general. For $K=3$ numerical integration
in Eq.(\ref{F2}) leads to $w_\phi \approx 71$.
In a similar way, we find critical values $w_r$ and $w_\phi$ for several other branching numbers $K$, the results are summarized in the Table 1.
\begin{table}[h!]
\centering
\renewcommand{\arraystretch}{1.5}
\begin{tabular}{|l||c|c|c|c|}
\hline
$K$ & 3 &  4 & 5 & 6  \\ \hline
$w_r$ & $ 53,7 $ & $ 80,6 $ & $ 109 $ & $ 139 $   \\ \hline
$w_\phi$ &  $71 $  &  $100$ &  $125$  &  $167$ \\  
\hline
\end{tabular}

\end{table}

\subsubsection{Account for self-energy correction}

%\subsubsection{Relaxation rate threshold}

\vspace{0.5cm}
\hspace{1cm}\textit{1). Relaxation rate}
\vspace{0.5cm}

Much better analytical approximation was proposed for the  problem that is very similar to the one defined by Eq.(\ref{SGammar}),  in Ref.~\cite{AKI2018}. 
They show that self-energy effects 
may be accounted for by the replacement of bare distribution $P_0(h)$ by \textit{effective} distribution function $P_1(h)$
\begin{equation}
P_1(h) = \frac{1}{W - 4J^2/W} \theta (\frac{W}{2} - |h|)\theta(|h|- \frac{2J^2}{W})
\label{P1}
\end{equation}
which accounts for the absence of too small resonance denominators in the R.H.S. of Eq.(\ref{SGammar}), due to self-energy corrections. An important feature of the distribution (\ref{P1}) is its symmetry: $P_1(h/W) = P_1(W/h)$. Due to this symmetry,
one does not need  to optimize over the value of exponent $x$, 
as it was done in Eqs.~(\ref{crit}) above. It was shown already in Ref.~\cite{AAT} that the optimal exponent is now $x=1/2$.
Instead of two equations in Eq.~(\ref{crit}),  we put $x=1/2$
into the definition of function $F(x)$ in Eq.~(\ref{F1}) 
and use the second equation of Eqs.~(\ref{crit})
only: $F_1(\frac12) = 0$.
Then the critical value  $j_r = J_r/W$ can be found~\cite{AKI2018} from 
\begin{equation}
K J \int P_1(h)\frac{dh}{h} = 1
\label{Jr2}
\end{equation}
Using Eq.(\ref{P1}) we get then an algebraic equation 
\begin{equation}
 \frac{w_r}{2} - \frac{2}{w_r} = 2 K \ln\frac{w_r}{2}\, 
 \label{solution1}
\end{equation}
whose solution is $w_r \approx 33$ for $K=3$ in very good agreement with numerical results~\cite{AndNum2010} (agreement persists for other values of $K$ as well).

%\subsubsection{Dephasing rate threshold}

\vspace{0.5cm}
\hspace{1cm}\textit{2). Dephasing rate}
\vspace{0.5cm}

Main idea for account of $\Re\Sigma$ is the same as above:  we use $x=1/2$ and integrate over region on the plane  $(h_1,h_2)$ where $|\Re\Sigma| < W/2$ and also both $|h_{1,2}| < W/2$. 
Specifically, we need to solve
\begin{equation}
 J^2K^2\int dh_1 dh_2 P_2(h_1,h_2)\left|\frac{1}{h_1} + \frac{1}{h_2}\right|\frac{1}{|h_1-h_2|} = 1
\label{P12}
\end{equation}
with renormalized distribution 
$$P_2(h_1,h_2) = \frac{1+a}{W^2}\theta(\frac{W}{2}- |S|)\theta (\frac{W}{2} - |h_1|)\theta (\frac{W}{2} - |h_2|)
$$
 where
\begin{equation}
S \equiv \Re\Sigma = J^4 \left(\frac{1}{h_1} + \frac{1}{h_2}\right)^2\frac{1}{|h_1-h_2|}
\end{equation}
and $ a \ll 1$ accounts for the change of normalization of the distribution due to restriction $|s| < W/2$; this is
very small effect which can be neglected.

In dimensionless units  we need to solve for the value of $j$ the following
equation (main interest is in $K=3$):
\begin{equation}
 \int_{-1/2}^{1/2} \int_{-1/2}^{1/2} dx dy \left|\frac{1}{x} + \frac{1}{y}\right|
\frac{\theta(\frac12 - |s|)}{|x-y|} = \frac{w^2}{K^2}
\label{P13}
\end{equation}
where 
$$
s = w^{-4}\left(\frac{1}{x} + \frac{1}{y}\right)^2\frac{1}{|x-y|}
$$
Numerical solution for $K=3$ leads to $w_\phi \approx 50$. Ratio $w_\phi/w_r = 1.5$.  
In a similar way, we find critical values $w_r$ and $w_\phi$ for several other branching numbers $K$, the results are summarized in the Table 2.
\begin{table}[h!]
\centering
\renewcommand{\arraystretch}{1.5}
\begin{tabular}{|l||c|c|c|c|}
\hline
$K$ & 3 &  4 & 5 & 6  \\ \hline
$w_r$ & $33$ & $51$ & $71$ & $90$   \\ \hline
$w_\phi$ &  $50$  &  $77$ &  $100$  &  $133$ \\  
\hline
\end{tabular}

\end{table}

Results collected in Table 1 and Table 2 demonstrate that $w_\phi > w_r$ for all various values of $K$ and within two different approximation schemes.

\end{document}